\documentclass{SciPost}

\hypersetup{
    colorlinks,
    linkcolor={red!50!black},
    citecolor={blue!50!black},
    urlcolor={blue!80!black}
}

\usepackage{bm}
\usepackage{gensymb}
\usepackage[normalem]{ulem}
\usepackage{placeins}
\newcommand{\nn}{\nonumber \\}

\newcommand\redout{\bgroup\markoverwith{\textcolor{red}{\rule[.5ex]{2pt}{0.4pt}}}\ULon}

\newcommand{\blue}[1]{#1}
\newenvironment{revision}{}{}

\usepackage[bitstream-charter]{mathdesign}
\DeclareSymbolFont{usualmathcal}{OMS}{cmsy}{m}{n}
\DeclareSymbolFontAlphabet{\mathcal}{usualmathcal}

\fancypagestyle{SPstyle}{
\fancyhf{}
\lhead{\colorbox{scipostblue}{\bf \color{white} ~SciPost Physics }}
\rhead{{\bf \color{scipostdeepblue} ~SciPost Phys. 21, 109 (2026) }}

\fancyfoot[C]{\textbf{\thepage}}
}

\begin{document}

\pagestyle{SPstyle}

\begin{center}{\Large \textbf{\color{scipostdeepblue}{
%%%%%%%%%% TODO: Write your article's title here
Topological phase transition of deformed ${\mathbb Z}_3$ toric code\\
}}}\end{center}

\begin{center}\textbf{
% Use (full) first name (+ middle name initials) + surname format.
% Separate subsequent authors by a comma, omit comma and use "and" for the last author.
% Mark the corresponding author(s) with a superscript symbol in this order
% \star, \dagger, \ddagger, \circ, \S, \P, \parallel, ...
Yun-Tak Oh\textsuperscript{1} and
Hyun-Yong Lee\textsuperscript{1, 2$\star$}
}\end{center}

\begin{center}
{\bf 1} Division of Display and Semiconductor Physics, Korea University, Sejong 30019, Korea
\\
{\bf 2} Department of Applied Physics, Graduate School, Korea University, Sejong 30019, Korea
%%%%%%%%%% END TODO: AFFILIATIONS
% Provide email address of corresponding author(s)
\\[\baselineskip]
$\star$ \href{mailto:email1}{\small hyunyong@korea.ac.kr}
\end{center}

\section*{\color{scipostdeepblue}{Abstract}}
\textbf{\boldmath{%
{We investigate topological phase transitions in a family of
deformed $\mathbb Z_3$ toric-code wavefunctions prepared from a cluster state
by local deformations and projective measurements. Their norms map to the
$Q=3$ Potts model for single-parameter deformations and to a three-state
Ashkin--Teller-like (AT$_3$) construction with two independent four-spin
couplings in the general case. Projected entangled-pair-state (PEPS) and
variational uniform matrix-product-state (VUMPS) calculations identify the
toric-code (TC) phase and phases in which electric ($e$) anyons are confined or
condensed. These phases are separated by critical structures with central
charges $c=4/5$, $8/5$, and isolated $c=1$ antiferromagnetic (AFM) endpoints. A
normalized finite-distance $e$-anyon pair-state norm provides a
Fredenhagen--Marcu-type check of the confinement boundary, while the
topological data of the quantum double $D(\mathbb Z_3)$ imply a topological
entanglement entropy $\gamma=\log3$ throughout the gapped toric-code phase.
Relative to the
$\mathbb Z_2$ case, the absence of sign-change folding leaves the AFM endpoints
unfolded, and the extreme deformation reaches square ice with an emergent
$U(1)$ one-form symmetry, Hilbert-space fragmentation, and exact scar
configurations.}
}}

\vspace{\baselineskip}

%%%%%%%%%% BLOCK: Copyright information
% This block will be filled during the proof stage, and finilized just before publication.
% It exists here only as a placeholder, and should not be modified by authors.
\noindent\textcolor{white!90!black}{%
\fbox{\parbox{0.975\linewidth}{%
\textcolor{white!40!black}{\begin{tabular}{lr}%
  \begin{minipage}{0.50\textwidth}%
    {\small Copyright Y.-T. Oh and H.-Y. Lee. \newline
    This work is licensed under the Creative Commons \newline
    \href{https://creativecommons.org/licenses/by/4.0/}{Attribution 4.0 International License}. \newline
    Published by the SciPost Foundation.}
  \end{minipage} & \begin{minipage}{0.41\textwidth}
    {\small Received 2026-03-11 \newline Accepted 2026-09-21 \newline Published 2026-11-02 \newline
    {\footnotesize\doi{10.21468/SciPostPhys.21.5.109}}}%
  \end{minipage}
\end{tabular}}
}}
}
%%%%%%%%%% BLOCK: Copyright information

% For convenience during refereeing we turn on line numbers:
% \linenumbers

\vspace{10pt}
\noindent\rule{\textwidth}{1pt}
\tableofcontents
\noindent\rule{\textwidth}{1pt}
\vspace{10pt}

% Write your article contents here, starting from first \section.
% An example structure is given below.

\section{Introduction}
\label{sec:intro}

Cluster states have garnered significant attention in quantum computation and information due to their role as universal resources for measurement-based quantum computation (MBQC) \cite{raussendorf01prl01, raussendorf03pra, nielsen06repmat}. 
In contrast to the traditional circuit model, where unitary operations are applied sequentially, MBQC utilizes a highly entangled initial state\blue{---}the cluster state\blue{---}on which adaptive single-qubit measurements are performed to drive the computation \cite{briegel09natphys}. 
This paradigm shift has been extensively studied, both theoretically and experimentally, highlighting the versatility and scalability of cluster states in quantum information processing \cite{gross07prl, nest07prl}.

A particularly intriguing application of cluster states is their connection to topological quantum codes.
Recently, significant progress has been made in realizing topologically ordered states on programmable quantum simulators \cite{raussendorf05pra, brown11}. 
This approach offers an alternative method for constructing topologically ordered states within the MBQC framework, where logical operations are realized via carefully designed measurement protocols. 
The toric code \blue{(TC)}, introduced by Kitaev, serves as a foundational model in topological quantum error correction, characterized by its intrinsic fault tolerance and anyonic excitations \cite{kitaev03}. 
The interplay between MBQC and topological order underscores a profound link between quantum computation and condensed matter physics, motivating further exploration of cluster-state-based constructions of topological codes \cite{raussendorf07prl}.

Building upon this foundation, we extend the $\mathbb{Z}_2$ cluster state framework to a $\mathbb{Z}_3$ cluster state and demonstrate that a $\mathbb{Z}_3$ toric code state can be achieved through a similar measurement-based process. 
This generalization is particularly compelling as higher-dimensional qudits, such as qutrits, have garnered interest for their potential to enhance computational power and robustness against certain errors. 
Moreover, by modifying the $\mathbb{Z}_3$ cluster state prior to measurement, we can obtain a deformed $\mathbb{Z}_3$ toric code, which interpolates between different topological phases. 
Previous studies have extensively examined the phase diagram of the deformed $\mathbb{Z}_2$ toric code, revealing insights into topological phase transitions \cite{zhu-19prl}. 
In this work, we extend this analysis to the $\mathbb{Z}_3$ case, providing a systematic exploration of its phase diagram and the role of deformations in its topological properties.

\begin{revision}
The resulting diagram is a phase diagram of a measurement-prepared family of
wavefunctions, obtained from an exact two-dimensional norm mapping, rather
than the thermodynamic phase diagram of a $(2+1)$-dimensional toric-code
Hamiltonian in external fields. This distinction fixes the scope of the
critical theories and diagnostics discussed below.
\end{revision}

\begin{revision}
The $\mathbb Z_3$ problem also differs qualitatively from its $\mathbb Z_2$
counterpart. The clock algebra does not provide the qubit sign-change folding;
the two four-spin constraints that coincide modulo two become independent
modulo three; the unfolded phase diagram contains isolated $c=1$
antiferromagnetic (AFM)
endpoints; and the square-ice limit supports an emergent $U(1)$ one-form
symmetry with the exact scar count $2^{L+2}-4$. We foreground these structural
differences below rather than treating the qutrit case as a parameter-by-parameter
extension of the qubit construction.
\end{revision}

The remainder of this paper is structured as follows. In Sec.~\ref{sec:tc}, we introduce the formalism of the $\mathbb{Z}_3$ toric code and its relation to cluster states, outlining the measurement-based approach to generating topological states.
In Sec.~\ref{sec:deformed}, we describe how a deformed $\mathbb{Z}_3$ toric code state can be constructed by modifying the $\mathbb{Z}_3$ cluster state and discuss its corresponding parent Hamiltonian.
In Sec.~\ref{sec:phase}, we analyze the phase diagram of the deformed $\mathbb{Z}_3$ toric code by considering different deformation parameters and their impact on topological order.
In Sec.~\ref{sec:tn}, we employ tensor network techniques to further investigate the nature of the deformed $\mathbb{Z}_3$ toric code, including its connection to the Ashkin--Teller-like classical model and its projected entangled pair state (PEPS) representation.
Finally, in Sec.~\ref{sec:conclusion}, we summarize our findings and discuss potential future directions.

\section{Models}
\subsection{$\mathbb{Z}_N$ Toric Code}
\label{sec:tc}

\begin{figure}[b!]
    \centering
    \setlength{\abovecaptionskip}{5pt}
    \includegraphics[width=0.8\textwidth]{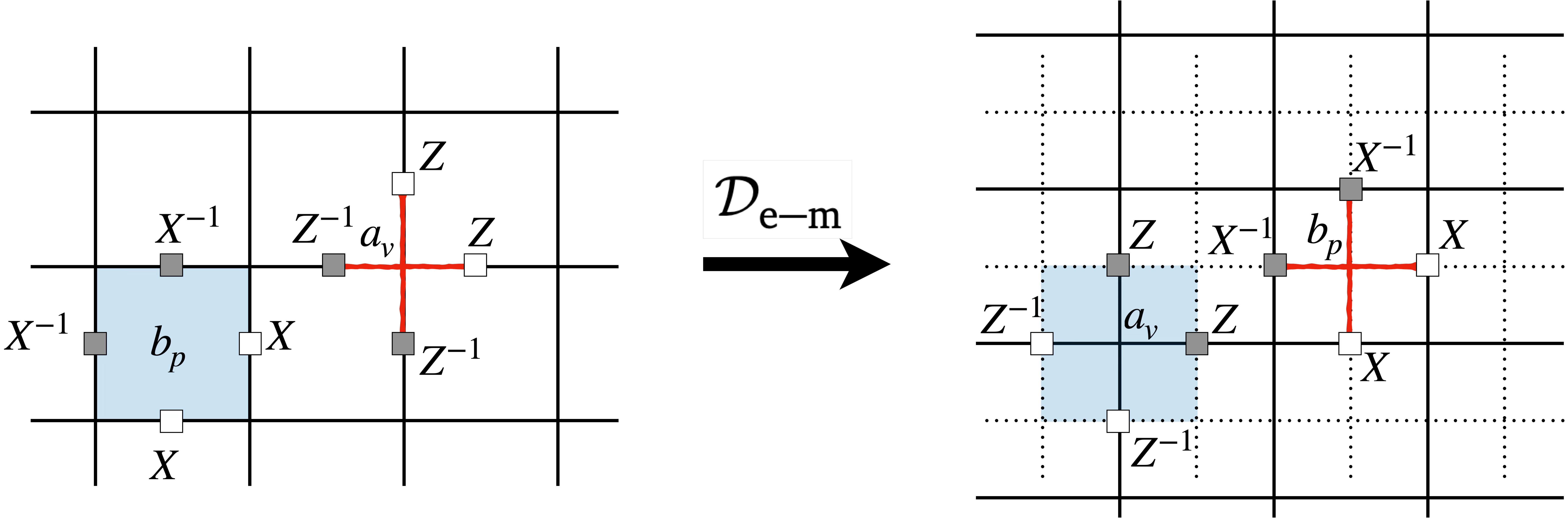}
    \caption{
    (Left panel) Vertex and plaquette stabilizers, $a_v$ and $b_p$, in the toric code model. (Right panel) The lattice and stabilizers following the dual transformation $\mathcal{D}_{e-m}$. The original lattice is shown with dotted lines, while the dualized lattice is depicted with solid lines.
    }
\label{fig:tc-model}
\end{figure}

The toric code Hamiltonian is given by:
\begin{align}
    H_{\rm TC} = - \sum_v A_v - \sum_p B_p,
\end{align}
where $v$ and $p$ denote the vertices and plaquettes of the two-dimensional square lattice. 
The projection operators are defined as:
\begin{align}
    A_v = \sum_{n=0}^{N-1} \left(a_v\right)^n, \quad
    B_p = \sum_{n=0}^{N-1} \left(b_p\right)^n,
\end{align}
where the vertex and plaquette operators
\begin{align}
    a_v = \prod_{l \in v} Z_l^{\xi_l}, \quad
    b_p = \prod_{l \in p} X_l^{\zeta_l}.
    \label{eq:stabilizer}
\end{align}
Here, the signs $\xi_l = \pm1$ and $\zeta_l = \pm1$ are defined as illustrated in Fig. \ref{fig:tc-model}, and $X_l$ and $Z_l$ are the $\mathbb{Z}_N$ generalized Pauli matrices defined on link $l$, satisfying $Z_lX_l = \omega X_lZ_l$ with $\omega^N = 1$. These matrices have eigenvalues $1, \omega, \cdots, \omega^{N-1}$.
It is straightforward to verify that the projection operators commute with each other, i.e., \([A_v, A_{v'}] = 0\), \([B_p, B_{p'}] = 0\), and \([A_v, B_p] = 0\). Consequently, the ground state is stabilized by the conditions \(a_v = 1\) for all \(v\) and \(b_p = 1\) for all \(p\). An electric anyon excitation with charge $a$ at $v$ corresponds to $a_v = \omega^a$, while a magnetic anyon excitation with charge $b$ at $p$ corresponds to $b_p = \omega^b$, where $a, b = 0, \cdots, N-1$. Notably, the model exhibits the electric-magnetic duality~\cite{zhu-19prl}:
\begin{align}
    {\cal D}_{\rm e-m}: ~~ X \leftrightarrow Z,
    \quad \text{and} \quad 
    v \leftrightarrow p.
\label{eq:toric-code-dual}
\end{align}

On a torus, the model is characterized by two $X$ holonomy operators defined as 
\begin{align}
W_X^x = \prod_{l \in {\cal C}_x } X_l, 
\quad \text{and} \quad
W_X^y = \prod_{l \in {\cal C}_y } X_l, 
\end{align}
where ${\cal C}_x$ and ${\cal C}_y$ represent non-contractible paths on the lattice that wind around the system in the $x$ and $y$ directions, respectively. Similarly, the model features two $Z$ holonomies given by
\begin{align}
W_Z^x = \prod_{l \in {\cal C}_x' } Z_l, 
\quad \text{and} \quad
W_Z^y = \prod_{l \in {\cal C}_y' } Z_l,
\end{align}
with ${\cal C}_x'$ and ${\cal C}_y'$ denoting non-contractible paths on the dual lattice winding in the $x$ and $y$ directions, respectively. These holonomy operators satisfy the following Heisenberg algebra:
\begin{align}
W_Z^y  W_X^x = \omega  W_X^x W_Z^y, \quad \text{and} \quad W_Z^x  W_X^y = \omega  W_X^y W_Z^x.
\end{align}
Due to these Heisenberg commutation relations, the TC model exhibits a ground state degeneracy of $N^2$.

% One of the degenerate ground states can be constructed by applying the projection operator to a reference product state: 
% \begin{align}
% |{\rm TC}_N\rangle 
% = \prod_p B_p |{\bm 0}\rangle.
% \label{eq:toric-code-gs}
% \end{align}
% Here, $|{\bm 0}\rangle = \otimes_i |0\rangle$, and $|0\rangle$ is the local qudit state satisfying $Z|0\rangle = |0\rangle$. This state is an eigenstate of both the $W_Z^x$ and $W_Z^y$ holonomies, each with eigenvalue 1. The other ground states, in the basis of the $W_X^y$ and $W_X^x$ holonomies, can be expressed as:
% %
% \begin{align}
% |{\rm TC}_N\rangle 
% = \sum_{i,j=0}^{N-1} \left(W_X^y\right)^i \left(W_X^x\right)^j \prod_p B_p |{\bm 0}\rangle,
% \label{eq:toric-code-gs-2}
% \end{align}
% %
% which is an eigenstate of the $W_X^y$ and $W_X^x$ holonomies, each with eigenvalue 1. In this work, we occasionally use either expression in Eqs.\,\eqref{eq:toric-code-gs} or \eqref{eq:toric-code-gs-2} in argument and focus on exploring the quantum phases and their associated topological properties that arise through its deformation.. The choice of which basis to use not significantly affect the overall argument. 

One of the degenerate ground states can be constructed by applying the projection operator to a reference product state, as follows: 
\begin{align}
|{\rm TC}_N\rangle 
= \prod_p B_p |{\bm 0}\rangle.
\label{eq:toric-code-gs}
\end{align}
Here, \( |{\bm 0}\rangle = \otimes_i |0\rangle \), where \( |0\rangle \) is the local qudit state satisfying \( Z|0\rangle = |0\rangle \). This state is an eigenstate of both the \( W_Z^x \) and \( W_Z^y \) holonomies, each with eigenvalue 1. The other ground states, expressed in the basis of the \( W_X^y \) and \( W_X^x \) holonomies, are given by:
\begin{align}
|{\rm TC}_N\rangle 
= \sum_{i,j=0}^{N-1} \left(W_X^y\right)^i \left(W_X^x\right)^j \prod_p B_p |{\bm 0}\rangle,
\label{eq:toric-code-gs-2}
\end{align}
which is an eigenstate of the \( W_X^y \) and \( W_X^x \) holonomies, each with eigenvalue 1. 

In this work, we use either expression from Eqs.\,\eqref{eq:toric-code-gs} or \eqref{eq:toric-code-gs-2} interchangeably, depending on the context, and focus on analyzing the quantum phases and their associated topological properties that emerge through deformation. The choice of basis does not significantly impact the overall argument.

% On a torus, the model exhibits  the $N^2$ ground state degeneracy, \textcolor{red}{with the degenerate states} distinguished by topological Wegner-Wilson loop operators. \hy{No need to explain about Wegner-Wilson loop operator? If so, ignore this.} \oh{The Wegner-Wilson loop may need to be introduced in later section. But, I am not sure where to put this. We should discuss about this.}

% \begin{align}
% |{\rm TC}_N\rangle 
% %= \prod_{p} \left( \mathbb{I}_{p} + b_p + \cdots + \left( b_p\right)^{N-1}\right) |{\bm 0}\rangle 
% = \prod_p B_p |{\bm 0}\rangle,
% \label{eq:toric-code-gs}
% \end{align}
% %

% In this study, we begin with the following specific ground state and focus on analyzing the emerging quantum phases and their topological properties resulting from its deformation:
% \redout{In this study, we focus on one specific ground state defined as follows:}
% It should be noted that our analysis in this study remains unaffected by the choice of the ground state.

\subsection{Cluster State: Toric Code Ground State}
\label{sec:cluster}

\begin{figure}[t!]
\centering
\setlength{\abovecaptionskip}{5pt}
\includegraphics[width=0.9\textwidth]{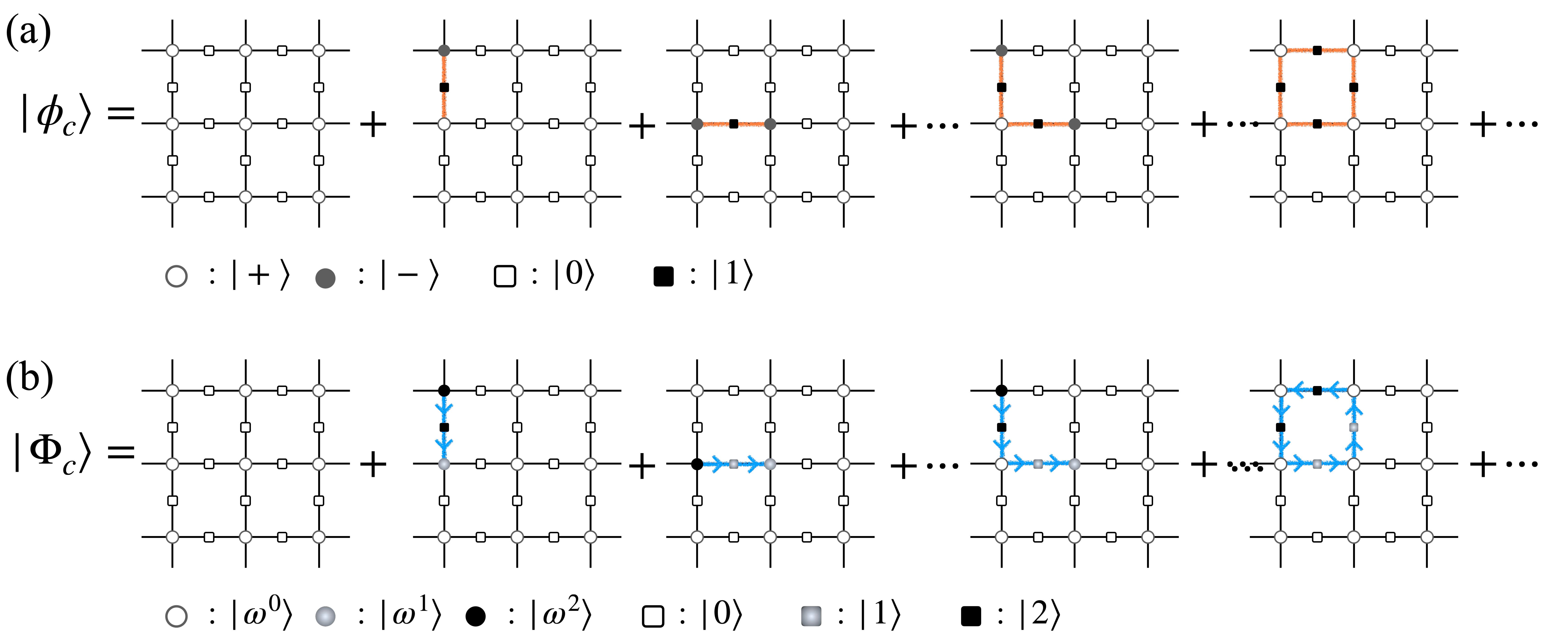}
\caption{Expansion of the cluster state on the Lieb lattice for (a) the $\mathbb{Z}_2$ case and (b) the $\mathbb{Z}_3$ case. Each term in the product of Eq.\,\eqref{eq:z2-cluster} (or Eq.\,\eqref{eq:z3-cluster}) corresponds to a dimer configuration. In (a), empty circles and squares represent $|+\rangle$ and $|0\rangle$, while filled ones represent $|-\rangle$ and $|1\rangle$; orange lines denote links where $Z_{v_l}X_l Z_{v_l'}$ is applied. In (b), circles and squares with three shadings represent the $\mathbb{Z}_3$ eigenstates $|\omega^0\rangle$, $|\omega^1\rangle$, $|\omega^2\rangle$ and $|0\rangle$, $|1\rangle$, $|2\rangle$, respectively; blue arrows indicate the direction of the $ZXZ$ dimers.
} \label{fig:z2-cluster}
\end{figure}

The cluster state serves as a significant resource in quantum information and topological matter~\cite{raussendorf01prl, nielsen06repmat}. In particular, it has been demonstrated that the TC state can be obtained by applying appropriate measurements to the cluster state~\cite{raussendorf05pra, brown11}. Here, we discuss how to generate the $\mathbb{Z}_2$ TC state from the cluster state through specific measurement protocols and extend this approach to generalize the generation of the $\mathbb{Z}_3$ TC state.

We begin with the $\mathbb{Z}_2$ cluster state defined on Lieb lattice as follows:
\begin{align}
|\phi_2 \rangle = \left[\prod_{l} \left(I + Z_{v_l} X_l Z_{v_l'} \right)  \right] | \bm{+}\rangle^{\otimes v} | \bm{0} \rangle^{\otimes l},
% |\phi \rangle = \prod_{l} \left(I + X_v Z_l X_{v'} \right) | \bm{0}_v\rangle \otimes | \bm{+}_l \rangle,
\label{eq:z2-cluster}
\end{align}
where $v_l$ and $v_l'$ are neighboring vertices connected by the link $l$, $|\bm{0}\rangle^{\otimes l} \equiv \otimes_l |0_l\rangle $, and $|\bm{+}\rangle^{\otimes v} \equiv \otimes_v |+_v\rangle $ with $|+_v\rangle$ being an eigenstate of $X_v$. In a quantum circuit, the cluster state can be prepared by applying controlled-Z (CZ) gates to neighboring qubits residing on the Lieb lattice~\cite{nielsen06repmat}.

Interpreting \( Z_{v_l} X_l Z_{v_l'} \) as a dimer on the link \( l \), the expansion of the product over \( l \) in Eq.\,\eqref{eq:z2-cluster} results in an equal-weight superposition of all possible dimer configurations, including those that allow vertices to be shared, as illustrated in Fig.\,\ref{fig:z2-cluster}\,(a); empty circles and squares represent $|+\rangle$ and $|0\rangle$, respectively, while filled circles and squares represent the states $|-\rangle$ and $|1\rangle$. The orange lines denote the links where the operator $Z_{v_l} X_l Z_{v_l'}$ has been applied.
% \redout{Expanding the product over $l$ in Eq.\,\eqref{eq:z2-cluster} yields an \red{equal-weight} superposition of the following states: the vacuum state \red{$| \bm{+}\rangle^{\otimes v} | \bm{0} \rangle^{\otimes l}$}, all states obtained by applying $Z_{v_l} X_l Z_{v_l'}$ on a single link of the vacuum state, all states obtained by applying $Z_{v_l} X_l Z_{v_l'}$ on two links of the vacuum state, and so on. This expansion is illustrated in Fig.\,\ref{fig:z2-cluster}(a);} 

% \redout{When the orange lines are connected, the spins at the junctions of the connected lines are restored to $|+\rangle$, but the spins on the links occupied by the orange lines remain in the flipped state $|1\rangle$. As a result, when the orange lines form a closed loop, the spins on the vertices crossed by the orange loop become $|+\rangle$, whereas the spins on the links occupied by the loop become $|1\rangle$.}
% \redout{eliminate all configurations containing open loops of orange lines. The remaining states correspond to all configurations containing closed loops of orange lines in any form.} 
The operation of \( Z_{v_l} X_l Z_{v_l'} \) flips the local qubits in its support as \( |0\rangle \leftrightarrow |1\rangle \) and \( |+\rangle \leftrightarrow |-\rangle \). Therefore, the qubit state on the link \( l \) occupied by the dimer is \( |1_l\rangle \), while the one on the vertex \( v \) shared by an odd\,(even) number of dimers is \( |-_v\rangle \) (\( |+_v\rangle \)).
As a result, the states with the dimers forming closed loops do not contain the $|-\rangle$ state on any vertex.
By performing a forced measurement on the vertex qubits in the \( |+\rangle \) state, only configurations with closed loops remain, where the link qubits on the loops are in the \( |1\rangle \) state, while all other link qubits are in the \( |0\rangle \) state.
This resulting state corresponds to the ground state of the $\mathbb{Z}_2$ toric code:
\begin{align}
|{\rm TC}_2\rangle \propto \left(\langle + |\right)^{\otimes v}| \phi_2 \rangle
= \sum_{c_i} |c_i\rangle,
\end{align}
which is equivalent to the state in Eq.\,\eqref{eq:toric-code-gs-2}.
Here, the summation runs over all possible closed-loop configurations $c_i$ of $|1\rangle$ states in the background of $|0\rangle$ state. 

% \redout{The process of generating the TC state through measurements on the cluster state can be extended to $\mathbb{Z}_N$ systems.} 
% The protocol can be naturally extended to the \(\mathbb{Z}_N\) toric code (TC) state using the \(\mathbb{Z}_N\) cluster state and suitable measurement schemes. In particular, for the \(\mathbb{Z}_3\) case, the cluster state is defined as:
% %
% \begin{align}
% |\phi_3 \rangle = \left[\prod_{l} \left(I + Z_{v_l}^\dag X_l Z_{v_l'} + Z_{v_l} X_l^\dag Z_{v_l'}^\dag \right) \right]
% | \bm{\omega}^0\rangle^{\otimes v}  | \bm{0}\rangle^{\otimes l},
% \label{eq:z3-cluster}
% \end{align}
% %
% where \( |\bm{0}\rangle^{\otimes l} \equiv \otimes_l |0_l\rangle \), with \( |0\rangle \) being the eigenstate of the \(\mathbb{Z}_3\) operator \( Z_l \), and \( |\bm{\omega}^0\rangle^{\otimes v} \equiv \otimes_v |\omega^0_v\rangle \), and \( |\omega^0_v\rangle \) is the eigenstate of the \(\mathbb{Z}_3\) operator \( X_v \). Here, \(v_l\) refers to the vertex located to the left (or down) of \(l\), and \(v_l'\) refers to the vertex located to the right (or up) of \(l\).

The protocol can be naturally extended to the \(\mathbb{Z}_N\) TC state using the \(\mathbb{Z}_N\) cluster state and suitable measurement schemes. In particular, for the \(\mathbb{Z}_3\) case, the cluster state is defined as:
\begin{align}
|\phi_3 \rangle = \left[\prod_{l} \left(I + Z_{v_l}^\dag X_l Z_{v_l'} + Z_{v_l} X_l^\dag Z_{v_l'}^\dag \right) \right]
| \bm{\omega}^0\rangle^{\otimes v}  | \bm{0}\rangle^{\otimes l}.
\label{eq:z3-cluster}
\end{align}
Here, \( |\bm{0}\rangle^{\otimes l} \equiv \otimes_l |0_l\rangle \), with \( |0\rangle \) being the eigenstate of the \(\mathbb{Z}_3\) operator \( Z_l \), and \( |\bm{\omega}^0\rangle^{\otimes v} \equiv \otimes_v |\omega^0_v\rangle \), where \( |\omega^0_v\rangle \) is the eigenstate of the \(\mathbb{Z}_3\) operator \( X_v \). Note that \(v_l\) refers to the vertex located left (down) from \(l\), and \(v_l'\) refers to the vertex located right (up) from \(l\).

% The ordering of $v$, $l$, and $v'$ is specified such that they follow the positive direction of the $x$ or $y$ coordinate.\hy{I do not understand how to define the direction here.}

% \redout{As in the $\mathbb{Z}_2$ case, expanding the product over \(l\) in Eq.\,\eqref{eq:z3-cluster} generates a superposition of all possible open and closed loop configurations, as illustrated in Fig.\,\ref{fig:z2-cluster}(b). }
% \redout{Unlike the \(\mathbb{Z}_2\) case, the links in the \(\mathbb{Z}_3\) configuration acquire directionality: links where \( Z_{v_l}^\dag X_l Z_{v_l'} \) is applied are marked with rightward (or upward) lines, while links where \( Z_{v_l} X_l^\dag Z_{v_l'}^\dag \) is applied are marked with leftward (or downward) lines.} \redout{Specifically, the local spins on vertices} 

Similar to the \(\mathbb{Z}_2\) case, we consider the \(ZXZ\) operation as a dimer, but now there are two distinct types. As will be shown, it is useful to distinguish these dimers by assigning them a direction: the rightward (upward) dimer \(Z_{v_l}^\dag X_l Z_{v_l'}\), and the leftward (downward) dimer \(Z_{v_l} X_l^\dag Z_{v_l'}^\dag\). Consequently, expanding the product over \(l\) in Eq.\,\eqref{eq:z3-cluster} generates a superposition of all possible two-species dimer configurations as depicted in Fig.\,\ref{fig:z2-cluster}\,(b). Here, the empty circles and squares correspond to \( |\omega^0\rangle \) and \( |0\rangle \), respectively, while the circles and squares filled in gray represent \( |\omega^1\rangle \) and \( |1\rangle \), and those filled in black represent \( |\omega^2\rangle \) and \( |2\rangle \). 
Then, the vertex qutrits take the state \( |\omega^{n_v} \rangle \), where \( n_v \equiv n^{\rm in}_v - n^{\rm out}_v \), with \( n^{\rm in}_v \) and \( n^{\rm out}_v \) denoting the number of dimers incoming to and outgoing from the vertex, respectively. 

Performing a forced measurement on vertices in the state $|\omega^0\rangle_v$ leaves only the closed $\mathbb{Z}_3$ loop configurations, corresponding to configurations where $n_v = 0$ for all vertices.
The resulting state can be interpreted as an equal-weight superposition of all possible domain-wall configurations of the 3-state Potts model, where the link qutrits are in $|1\rangle$ or $|2\rangle$, depending on the direction of the domain wall, against a background of $|0\rangle$. Indeed, this corresponds to the $\mathbb{Z}_3$ TC state:
\begin{align}
|{\rm TC}_3\rangle \propto  \left(\langle \omega^0 |\right)^{\otimes v} | \phi_3 \rangle = \sum_{c_i} |c_i\rangle,
\label{eq:z3tc-cluster-gs}
\end{align}
where $|c_i \rangle$ denotes a certain domain-wall configuration.
In each configuration, the states on links unoccupied by any loops correspond to $|0\rangle$, the states on links with direction aligned with the positive coordinate axis correspond to $|1\rangle$, and the states on links with direction aligned with the negative coordinate axis correspond to $|2\rangle$.

\section{Generating Deformed $\mathbb{Z}_3$ Toric Code}
\label{sec:deformed}
In this section, we describe the deformation protocol applied to the \(\mathbb{Z}_3\) cluster state, which, following the same measurement protocol discussed in the previous section, results in a deformed \(\mathbb{Z}_3\) TC state. Then, we recast the deformed TC state into an alternative basis, enabling a more straightforward and insightful analysis. Finally, we derive the parent Hamiltonian that hosts the deformed \(\mathbb{Z}_3\) TC states as its ground states.

\subsection{Deformed $\mathbb{Z}_3$ Cluster State }
% \redout{By modifying Eq.\,\eqref{eq:z3-cluster} and performing successive measurements, one can obtain the deformed $\mathbb{Z}_3$ toric code state. Consider a system defined on the Lieb lattice with local $\mathbb{Z}_3$ spins. First, we prepare a local state $|\theta^0\rangle$ on each link, defined as:}
We begin with defining a rotated qutrit state as follows:
\begin{align}
|0\rangle \rightarrow 
|\theta^0\rangle \equiv \cos \theta |0\rangle + \frac{\sin\theta}{\sqrt{2}} \left( |1\rangle + |2\rangle \right).
\label{eq:z3-cluster-theta-0}
\end{align}
Here, $|0\rangle$, $|1\rangle$, and $|2\rangle$ are eigenstates of the $\mathbb{Z}_3$ operator $Z$ with eigenvalues $1$, $\omega$, and $\omega^2$, respectively, where $\omega^3 = 1$. 
The state $|\theta_l^0\rangle$ can be prepared by applying an appropriate unitary operator to $|0\rangle$.
%
% These eigenstates are represented as:
% %
% \begin{align}
% |\omega^0\rangle
% \propto
% \begin{pmatrix}
% 1 \\ 1 \\ 1
% \end{pmatrix},\quad
% %
% |\omega^1\rangle
% \propto
% \begin{pmatrix}
% 1 \\ \omega \\ \omega^2
% \end{pmatrix},\quad
% %
% |\omega^2\rangle
% \propto
% \begin{pmatrix}
% 1 \\ \omega^2 \\ \omega
% \end{pmatrix}.
% \end{align}
We consider the above \( |\theta^0\rangle \) state as one of the basis states. To preserve the algebra, i.e., ensuring that the operator \( X \) permutes the basis states, we define the remaining two basis states as follows:
% \redout{The $\mathbb{Z}_3$ operator $X$ cyclically permutes the state $|\theta^0\rangle$ with the states $|\theta^1\rangle$ and $|\theta^2\rangle$ as follows:}
%
\begin{align}
|\theta^1\rangle &\equiv X |\theta^0\rangle =  \cos \theta |1\rangle + \frac{\sin\theta}{\sqrt{2}} \left( |2\rangle + |0\rangle \right), \nn 
|\theta^2\rangle &\equiv X^2 |\theta^0\rangle =  \cos \theta |2\rangle + \frac{\sin\theta}{\sqrt{2}} \left( |0\rangle + |1\rangle \right).
\label{eq:z3-cluster-theta-12}
\end{align}
\begin{figure}[b!]
    \centering
    \setlength{\abovecaptionskip}{5pt}
    \includegraphics[width=0.8\textwidth]{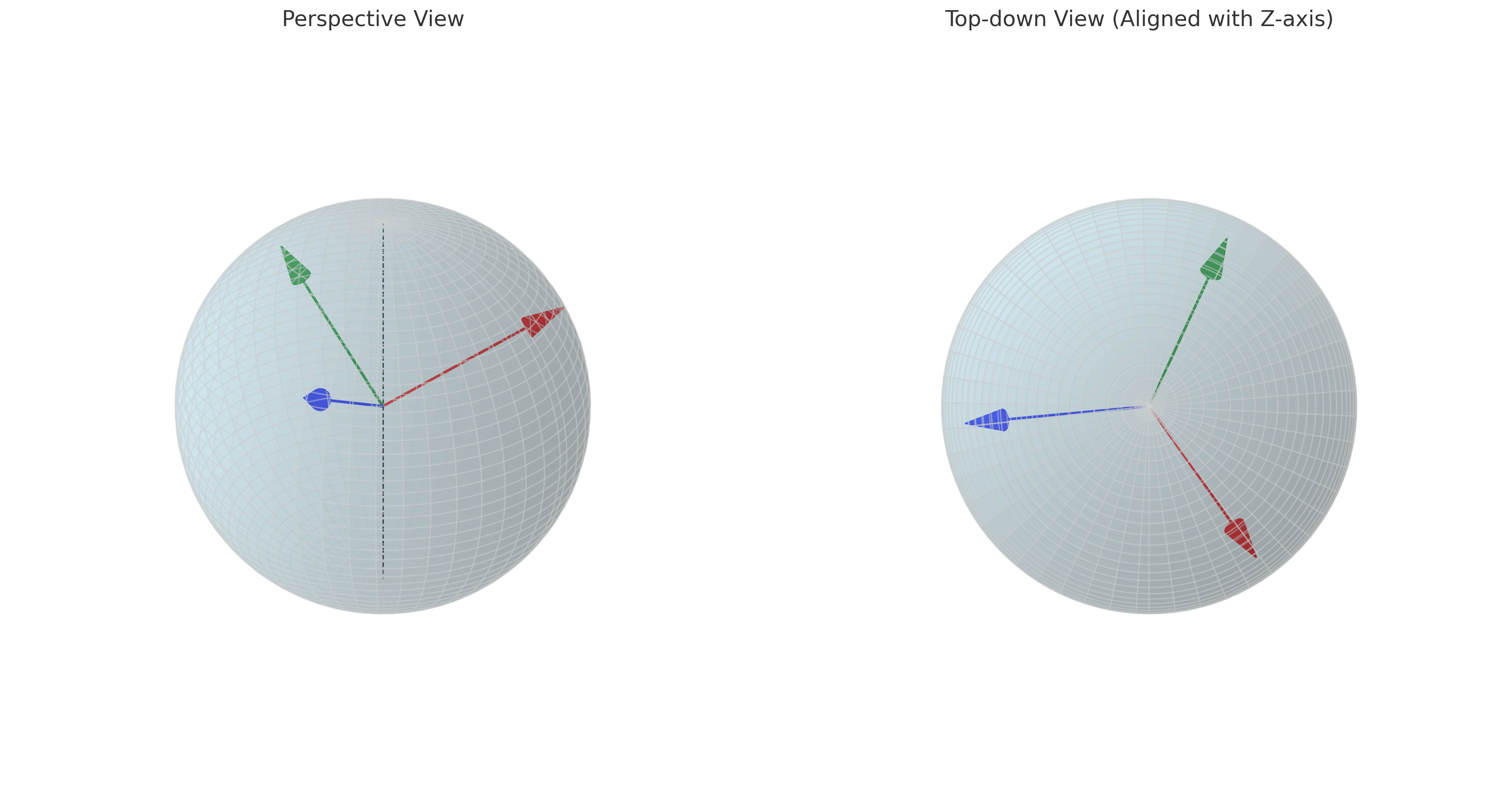}
    \caption{Visualization of the rotated basis states $|\theta^0\rangle$, $|\theta^1\rangle$, and $|\theta^2\rangle$ as vectors in the three-dimensional vector space. (Left) Perspective view showing the three vectors tilted at an angle $\theta$ from the $z$-axis. (Right) Top-down view along the $z$-axis, highlighting the $120^\circ$ rotational symmetry about the $[111]$-axis. As $\theta$ varies, the mutual overlap $\langle \theta^a | \theta^b \rangle$ interpolates between $-\frac{1}{2}$ (coplanar) and $1$ (collinear), with $0$ (orthonormal) at $\theta = 0$.}
\label{fig:rotated-basis}
\end{figure}
Each of the three states $|\theta^0\rangle$, $|\theta^1\rangle$, and $|\theta^2\rangle$ is normalized to $1$, but they are not mutually orthogonal:
\begin{align}
\langle \theta^a | \theta^b\rangle = \sin \theta \left( \frac{1}{2} \sin\theta + \sqrt{2} \cos\theta\right), \quad \text{for } a \neq b.
\end{align}
%
% \redout{The included angles between the three states are equivalent.} 
The basis states can be visualized as vectors in the three-dimensional vector space, each forming an angle of \( \theta \) with the \( z \)-axis and rotated by \( 120^\circ \) around the \([111]\)-axis relative to each other, as illustrated in Fig.\,\ref{fig:rotated-basis}.
Note that the overlap between the rotated basis states,  $\langle \theta^a | \theta^b\rangle$, varies from $-\frac{1}{2}$ to $1$ as $\theta$ changes. When $\langle \theta^a | \theta^b\rangle = 0$\,($\theta = 0$), the three states form an orthonormal basis spanning the three-dimensional Hilbert space. When \( \langle \theta^a | \theta^b \rangle = -\frac{1}{2} \)\,($\theta = \pi - \arctan\frac{1}{\sqrt{2}} \approx 144.7^\circ$), the three states lie within a two-dimensional Hilbert subspace (or a plane in the three-dimensional vector space). In contrast, when \( \langle \theta^a | \theta^b \rangle = 1 \)\,($\theta = \arctan\sqrt{2} \approx 54.7^\circ$), all three states become identical, collapsing into a single state.

% \redout{By replacing $| \bm{\omega}^0_v\rangle \otimes | \bm{0}_l \rangle$ in Eq.\,\eqref{eq:z3-cluster} with $| \bm{\omega}^0_v\rangle  \otimes |\bm{\theta}^0_l\rangle$, where $|\bm{\theta}^0_l\rangle = \prod_l |\theta^0_l\rangle$ is a tensor product state of $|\theta^0_l\rangle$ on every link, we define a cluster-like state:}
Now, our deformation protocol is straightforward: we smoothly rotate the link qutrits $|\bm{0}\rangle^{\otimes l} $ in the reference state into the rotated one $|\bm{\theta}^0\rangle^{\otimes l} \equiv \otimes_l |\theta^0_l\rangle $, resulting in the deformed cluster state:
\begin{align}
|\phi_3(\theta)\rangle  =  \prod_{l} \left(I + Z_{v_l}^\dag X_l Z_{v_l'} + Z_{v_l} X_l^\dag Z_{v_l'}^\dag  \right) 
| \bm{\omega}^0\rangle^{\otimes v}  |\bm{\theta}^0\rangle^{\otimes l}.
\label{eq:z3-cluster-like}
\end{align}
By performing a forced measurement on the vertex qutrits in the state $\langle \omega^0 |^{\otimes v}$, the resulting state is given as:
\begin{align}
|\Psi_3(\theta)\rangle 
= 
\left(\langle \omega^0 |\right)^{\otimes v}  |\phi_3(\theta)\rangle
=
\sum_{c_i} |\bar{c}_i\rangle.
\label{eq:z3tc-cluster-like}
\end{align}
Since the vertex qutrits remain unchanged before the measurement, the geometrical interpretation of $|\Psi_3(\theta)\rangle$ is preserved as the equal-weight superposition of domain-wall configurations of the 3-state Potts model. The difference is that $|0\rangle \rightarrow |\theta^0\rangle$ on unoccupied sites, and $|1\rangle \rightarrow |\theta^1\rangle$, $|2\rangle \rightarrow |\theta^2\rangle$ on loops, depending on the direction. It is noteworthy that two different loop configurations are no longer orthogonal.

\subsection{Deformed Toric Code State and Parent Hamiltonian}
The deformed \(\mathbb{Z}_3\) TC state\,$|\Psi_3(\theta)\rangle$, derived from the deformed cluster state, is equivalent to applying a filtering operation directly on the \(\mathbb{Z}_3\) TC state:
\begin{align}
    |\tilde{\Psi}(\beta_x)\rangle \equiv \prod_l e^{\frac{\beta_x}{2} \Gamma_l^x} |{\rm TC}_3\rangle,
    \quad \text{where} \quad 
    \Gamma^x = 
    \frac{1}{3}
    \begin{pmatrix}
    -1 & 2 & 2 \\
    2 & -1 & 2 \\
    2 & 2 & -1
    \end{pmatrix}.
\end{align}
%
% \redout{The wavefunction $|\Psi(\theta)\rangle$ in Eq.\,\eqref{eq:z3tc-cluster-like} can also be constructed through an alternative approach. Starting with the $\mathbb{Z}_3$ toric code ground state, a deformation operator is applied to transform the local basis states $|0\rangle$, $|1\rangle$, and $|2\rangle$ into $|\theta^0\rangle$, $|\theta^1\rangle$, and $|\theta^2\rangle$, respectively. The deformation operator is given by:}
%
% \begin{align}
% D_l(\beta_x) = \exp\left(\frac{\beta_x}{2} \Gamma^x_l\right)  \propto I_l + \tanh\left(\frac{\beta_x}{2}\right) \Gamma^x_l, 
% \quad \text{where} \quad 
% \Gamma^x = 
% \frac{1}{3}
% \begin{pmatrix}
% -1 & 2 & 2 \\
% 2 & -1 & 2 \\
% 2 & 2 & -1
% \end{pmatrix}.
% \end{align}
%
The local deformation operator $D_l(\beta_x) = \exp{\left(\frac{\beta_x}{2}\Gamma^x \right)}$ transforms the local basis states as follows:
% \redout{modifies the local basis states, disrupting their orthonormality and rotating them in the local basis. The resulting deformed states are:}
%
\begin{align}
|0\rangle \xrightarrow[]{D(\beta_x)} &
|\tilde{0}\rangle \propto 
\left(3 - \tanh\frac{\beta_x}{2}\right) |0\rangle 
+ 2 \tanh\frac{\beta_x}{2} \left(|1\rangle + |2\rangle\right), \nn 
|1\rangle \xrightarrow[]{D(\beta_x)} &
|\tilde{1}\rangle \propto 
\left(3 - \tanh\frac{\beta_x}{2}\right) |1\rangle 
+ 2 \tanh\frac{\beta_x}{2} \left(|0\rangle + |2\rangle\right), \nn 
|2\rangle \xrightarrow[]{D(\beta_x)} &
|\tilde{2}\rangle \propto 
\left(3 - \tanh\frac{\beta_x}{2}\right) |2\rangle 
+ 2 \tanh\frac{\beta_x}{2} \left(|0\rangle + |1\rangle\right).
\end{align}
Therefore, with proper normalization, one finds that $|\tilde{0}\rangle$, $|\tilde{1}\rangle$, and $|\tilde{2}\rangle$ are equivalent to $|\theta^0\rangle$, $|\theta^1\rangle$, and $|\theta^2\rangle$, as defined in Eqs.\,\eqref{eq:z3-cluster-theta-0} and \eqref{eq:z3-cluster-theta-12}. The overlap between any two states is given by:
\begin{align}
\langle \tilde{a} | \tilde{b} \rangle 
= \frac{4 \tanh\left(\frac{\beta_x}{2}\right)}
{3 \tanh^2\left(\frac{\beta_x}{2}\right) - 2 \tanh\left(\frac{\beta_x}{2}\right) + 3}.
\label{eq:beta-x-local-overlap}
\end{align}
As $\beta_x$ varies from $-\infty$ to $\infty$, the value of $\langle \tilde{a} | \tilde{b} \rangle$ changes from $-\frac{1}{2}$ to $1$. At $\langle \tilde{a} | \tilde{b} \rangle = 0$, the three states form an orthonormal basis spanning the three-dimensional space. When $\langle \tilde{a} | \tilde{b} \rangle = -\frac{1}{2}$, the states are restricted to a two-dimensional subspace, and when $\langle \tilde{a} | \tilde{b} \rangle = 1$, all three states collapse into a single vector.

The deformed states $|\tilde{0}\rangle$, $|\tilde{1}\rangle$, and $|\tilde{2}\rangle$ share characteristics with the states $|\theta^0\rangle$, $|\theta^1\rangle$, and $|\theta^2\rangle$ from Eqs.\,\eqref{eq:z3-cluster-theta-0} and \eqref{eq:z3-cluster-theta-12}, particularly in terms of the angles between them. By appropriately mapping the parameters $\beta_x$ and $\theta$, the states $|\theta^0\rangle$, $|\theta^1\rangle$, and $|\theta^2\rangle$ can be related to the deformed states $|\tilde{0}\rangle$, $|\tilde{1}\rangle$, and $|\tilde{2}\rangle$.

Thus, the wavefunction $|\Psi_3(\theta)\rangle$ in Eq.\,\eqref{eq:z3tc-cluster-like} is equivalent to the deformed $\mathbb{Z}_3$ toric code state, which we hereafter denote as $|\tilde{\Psi}(\beta_x)\rangle$:
\begin{align}
|\tilde{\Psi}(\beta_x)\rangle  
= \prod_l D_l(\beta_x) |{\rm TC}_3\rangle
= \sum_{c_i} \prod_l D_l(\beta_x) |c_i\rangle
= \sum_{c_i} |\tilde{c}_i\rangle.
\label{eq:z3tc-gs}
\end{align}

\begin{revision}
Here $D_l(\beta_x)$ is off-diagonal in the local $Z$ basis. It mixes
$|0\rangle$, $|1\rangle$, and $|2\rangle$ into the non-orthogonal states
$|\tilde 0\rangle$, $|\tilde 1\rangle$, and $|\tilde 2\rangle$, whose common
off-diagonal overlap is given in Eq.~\eqref{eq:beta-x-local-overlap}.
Consequently, $|\tilde c_i\rangle$ denotes a locally rotated configuration,
and distinct configurations need not remain orthogonal.
\end{revision}

Under the dual transformation defined in Eq.\,\eqref{eq:toric-code-dual}, the deformed toric code state transforms as:
\begin{align}
|\tilde{\Psi}(\beta_z)\rangle = \prod_l D_l(\beta_z) |{\rm TC}_3\rangle
= \sum_{c_i} |\tilde{c}_i\rangle,
\label{eq:z3-z-deformed}
\end{align}
where $\beta_x \rightarrow \beta_z$, and the deformation operator $D_l(\beta_z)$ is expressed as:
\begin{align}
D_l(\beta_z) = \exp\left(\frac{\beta_z}{2} \Gamma_l^z \right) \propto I_l + \tanh \left(\frac{\beta_z}{2}\right)\Gamma^z_l, 
\quad \text{where} \quad
\Gamma^z = U_{X\leftrightarrow Z} \Gamma^x U_{X\leftrightarrow Z}^\dag 
=
\begin{pmatrix}
1 & 0 & 0 \\
0 & -1 & 0 \\
0 & 0 & -1
\end{pmatrix}.
\end{align}
\begin{revision}
In contrast, $\Gamma^z$ is diagonal in the $Z$ basis, so
$D_l(\beta_z)$ only rescales each local basis state. Thus, in the $\beta_z$
representation, $|\tilde c_i\rangle$ is a scalar-weighted version of the
orthogonal bare configuration $|c_i\rangle$; its explicit configuration
weight is given below in Eq.~\eqref{eq:z3-weight}.
\end{revision}

Consequently, our deformation protocol can be interpreted as tuning the loop fugacity through the parameter \(\theta\) or \(\beta_z\), where the fugacity per link becomes \( e^{-\beta_z} \). This breaks the equal-weight configuration characteristic of the TC state, filtering a specific channel determined by the parameter \(\theta\) or \(\beta_z\).

\blue{Because $\Gamma^z$ is diagonal, this transformation is a rescaling rather
than a rotation.} Under the $\Gamma^z$ deformation, the three local states
$|0\rangle$, $|1\rangle$, and $|2\rangle$ in $|\tilde{c}\rangle$ transform as:
\begin{align}
|0\rangle \rightarrow 
|\tilde{0}\rangle  = |0\rangle, 
\quad
|1\rangle \rightarrow 
|\tilde{1}\rangle =
\exp\left(-\beta_z\right) |1\rangle,
\quad
|2\rangle \rightarrow
|\tilde{2}\rangle  = 
\exp\left(-\beta_z\right) |2\rangle.
\end{align}

\begin{revision}
For finite $\beta_z$, let $D=\prod_lD_l(\beta_z)$ and
$D_p=\prod_{l\in\partial p}D_l(\beta_z)$. It is convenient to shift the
undeformed stabilizer Hamiltonian by an irrelevant constant and work with the
local zero-energy constraints
\begin{align}
Q_v=3\mathbb I-A_v,
\qquad
Q_p=3\mathbb I-B_p.
\end{align}
They satisfy $Q_v|{\rm TC}_3\rangle=Q_p|{\rm TC}_3\rangle=0$.
Since the diagonal filter commutes with $A_v$, the vertex constraint is
unchanged. The filtered plaquette constraint is
\begin{align}
Q'_p
&=D_pQ_pD_p^{-1}
=2\mathbb I-C_{p,1}b_p-C_{p,2}b_p^2,
\label{eq:filtered-plaquette-constraint}
\end{align}
where the diagonal operators $C_{p,1}$ and $C_{p,2}$ are defined as follows:
\end{revision}
\begin{align}
 \includegraphics[width=0.5\textwidth]{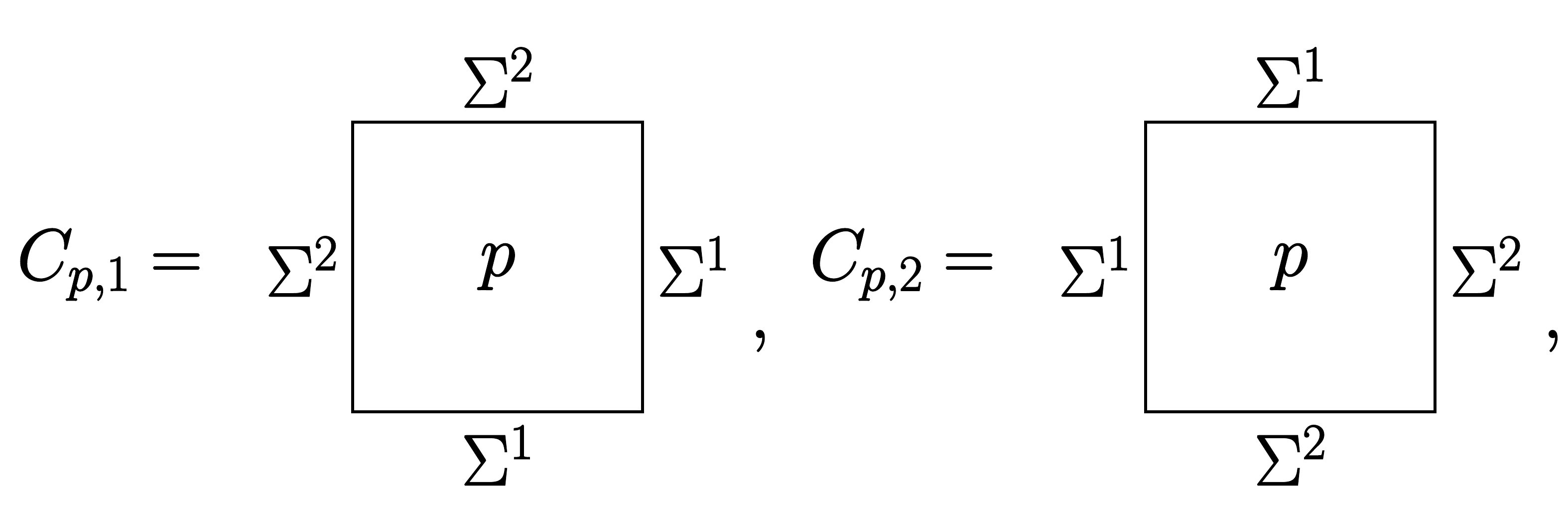}
\end{align}
where
\begin{align}
\Sigma^1 = 
\begin{pmatrix}
e^{\beta_z} & 0 & 0 \\
0 & e^{-\beta_z} & 0 \\
0 & 0 & 1
\end{pmatrix}, 
\quad
\Sigma^2 =
\begin{pmatrix}
e^{\beta_z} & 0 & 0 \\
0 & 1 & 0 \\
0 & 0 & e^{-\beta_z}  
\end{pmatrix}.
\end{align}

\begin{revision}
A direct single-link calculation gives
\begin{align}
D_lX_lD_l^{-1}=\Sigma_l^1X_l,
\qquad
D_lX_l^\dagger D_l^{-1}=\Sigma_l^2X_l^\dagger.
\end{align}
Multiplying these identities around a plaquette yields
\begin{align}
D_pb_pD_p^{-1}=C_{p,1}b_p,
\qquad
D_pb_p^2D_p^{-1}=C_{p,2}b_p^2,
\end{align}
and hence Eq.~\eqref{eq:filtered-plaquette-constraint}. The transformed
constraint annihilates the filtered state because
\begin{align*}
Q'_pD|{\rm TC}_3\rangle=DQ_p|{\rm TC}_3\rangle=0.
\end{align*}
The operators $C_{p,1}$ and $C_{p,2}$ therefore dress the plaquette flips;
they do not annihilate the state on their own.

A local Hermitian frustration-free parent Hamiltonian is
\begin{align}
H'_{\rm parent}
=\sum_vQ_v^\dagger Q_v+\sum_p{Q'_p}^\dagger Q'_p,
\label{eq:filtered-parent}
\end{align}
for which $|\tilde\Psi(\beta_z)\rangle=D|{\rm TC}_3\rangle$ is an exact
zero-energy ground state. The similarity transformation is used here only to
construct local annihilating constraints; it does not imply that this parent
Hamiltonian has the same excitation spectrum or thermodynamic transition as
the undeformed toric-code Hamiltonian. The singular limits
$\beta_z\to\pm\infty$ are understood by continuity from finite deformation.
\end{revision}

% ### Summary of the Section

One can summarize the contents of this section as follows:
\begin{itemize}
\item We introduced a cluster-like state $|\phi_3(\theta)\rangle$ constructed by applying a rotation operator to the vacuum configuration:
$$|{\bm +}_v\rangle \otimes |{\bm 0}_l\rangle \rightarrow |{\bm +}_v\rangle \otimes |{\bm \theta}^0_l\rangle.$$
\item By performing forced measurements on the vertex states, we obtained the state $|\Psi_3(\theta)\rangle$.
\item We showed that the same state can be equivalently constructed by applying the deformation operator $D_l(\beta_x)$ to the toric code wavefunction:
$$|\tilde{\Psi}(\beta_x)\rangle = \prod_l D_l(\beta_x) |{\rm TC}_3\rangle.$$
\item \blue{Through the dual transformation ${\cal D}_{\rm e-m}$, the
wavefunction is mapped to $|\tilde{\Psi}(\beta_z)\rangle$, and the dressed
local constraints and Hermitian frustration-free parent Hamiltonian
$H'_{\rm parent}$ are derived explicitly.}
\end{itemize}

\section{Phase Diagram of Deformed $\mathbb{Z}_3$ Toric Code}
\label{sec:phase}
\label{sec:phase-diagram}

\begin{revision}
We emphasize the object whose phases are studied below. The parameters
$\beta_x$ and $\beta_z$ define a family of filtered wavefunctions, and their
equal-time correlations are governed exactly by the two-dimensional classical
model obtained from the wavefunction norm. These transitions therefore belong
to the Rokhsar--Kivelson/conformal-quantum-critical wavefunction setting
~\cite{rokhsar88,ardonne04,castelnovo08}. They should not be confused with the
thermodynamic phase diagram of a toric-code Hamiltonian in transverse and
longitudinal fields, which is a $(2+1)$-dimensional quantum problem with a
three-dimensional classical gauge--Higgs description
~\cite{trebst07,vidal09,tupitsyn10}. The wavefunction problem is physically
relevant here because topological-code states can be prepared directly on
programmable quantum platforms~\cite{satzinger21,semeghini21,iqbal24}, including
a recent realization of the $\mathbb Z_3$ toric-code state on encoded qutrits
~\cite{iqbal25qutrit}.
\end{revision}

\subsection{Phase Diagram of $|\tilde{\Psi}(\beta_z)\rangle$}
\label{sec:beta-z-phase}
\subsubsection{Loop and Net Description}

\begin{figure}[ht]
\centering
\setlength{\abovecaptionskip}{5pt}
\includegraphics[width=0.9\textwidth]{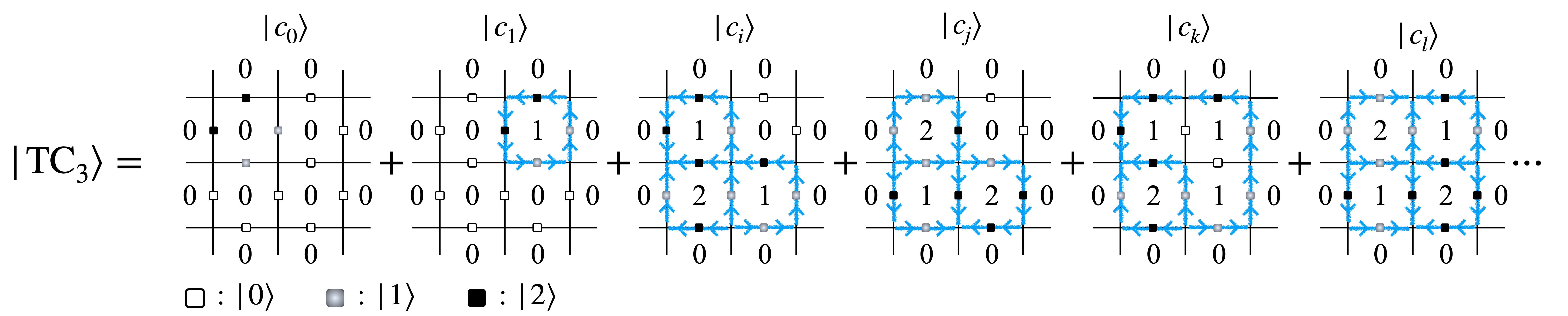}
\caption{Loop-gas representation of the $\mathbb{Z}_3$ toric code ground state $|{\rm TC}_3\rangle$ as an equal superposition of closed-loop configurations $|c_i\rangle$. The numbers at the center of each plaquette indicate how many times (mod 3) the operator $b_p$ has been applied. Blue arrows denote directed loops: links with rightward (downward) arrows carry $|1\rangle$, links with leftward (upward) arrows carry $|2\rangle$, and unoccupied links correspond to $|0\rangle$. Empty, gray, and black squares represent the local states $|0\rangle$, $|1\rangle$, and $|2\rangle$, respectively.}
\label{fig:z3-loop}
\end{figure}

To analyze the phase diagram of the deformed $\mathbb{Z}_3$ toric code (TC), the loop-gas configuration representation serves as an essential tool. As expressed in Eq.\,\eqref{eq:z3tc-gs}, the ground state of the $\mathbb{Z}_3$ TC can be written as an equal superposition of all possible closed-loop configurations:
\begin{align}
|{\rm TC}_3\rangle  = \sum_{c_i} |c_i\rangle.
\end{align}
These configurations are connected by the group $G$, which is generated by the operators $b_p$ and $b_p^2$. Starting from the fully magnetized configuration $|\bm{0}\rangle = \prod_l |0\rangle_l$, where $|0\rangle_l$ is the eigenstate of $Z_l$ with eigenvalue $+1$, all other configurations can be obtained by applying elements $g \in G$.

This representation is depicted in Fig.\,\ref{fig:z3-loop}, which highlights two distinct types of loops differentiated by directional arrows. The numbers at the center of each plaquette indicate how many times (mod 3) the $b_p$ operator has been applied to transition from $|\bm{0}\rangle$ to the given configuration. Loops form between neighboring plaquettes with differing numbers: a difference of $1\,{\rm mod}\,3$ produces arrows pointing right (or downward), while a difference of $2\,{\rm mod}\,3$ results in arrows pointing left (or upward). Links associated with rightward (or downward) arrows correspond to the local state $|1\rangle$, links with leftward (or upward) arrows correspond to $|2\rangle$, and links unoccupied by loops correspond to $|0\rangle$.

The charge of an $e$ anyon at a vertex $v$ is determined by the net divergence (mod 3) of the loops at that site. In the ground state, all configurations satisfy the divergence-free condition, ensuring that the total divergence at each vertex is zero (mod 3). $e$ anyons are created in pairs by cutting loops in the ground-state configurations, leading to two vertices with opposite $e$-anyon charges at the endpoints of the open loop.

More specifically, a pair of $e$ anyons can be created by flipping the local states along an open path ${\bm \gamma}$ that starts at vertex $i$ and ends at vertex $j$:
\begin{align}
\hat{O}_{\bm \gamma} =  \prod_{l \in {\bm \gamma}} X_l^{\xi_l},
\end{align}
where $\xi_l = \pm 1$ denotes the direction of the path ${\bm \gamma}$ on the link $l$. Applying $\hat{O}_{\bm \gamma}$ generates an $e$-anyon pair state:
\begin{align}
|e_i \bar{e}_j\rangle  = \hat{O}_{\bm \gamma} |{\rm TC}_3\rangle = \sum_{c_i} \hat{O}_{\bm \gamma} |c_i\rangle = \sum_{c_i} |c'_i\rangle.
\end{align}
In this expression, each configuration $|c'_i\rangle$ contains an open loop connecting vertices $i$ and $j$, whose specific form depends on the original configuration $|c_i\rangle$. The resulting set of configurations encompasses all possible open loops that connect $i$ and $j$.

At the toric code point, corresponding to the $\beta_z \rightarrow 0$ limit, the $e$ anyon pairs are deconfined. This can be verified by evaluating:
\begin{align}
\langle e_i \bar{e}_j | e_i \bar{e}_j \rangle = \sum_{c_i} \langle c'_i | c'_i \rangle.
\label{eq:z3-e-deconfine}
\end{align}
At the toric code point, all overlaps $\langle c'_i | c'_i \rangle$ between $e$-anyon pair configurations contribute equally to the summation. This equality arises because the closed loops in the original configurations $|c\rangle$ are scale-invariant, leading to uniform normalization across all configurations $|c'_i\rangle$. Consequently, the overlap $\langle e_i \bar{e}_j | e_i \bar{e}_j \rangle$ is independent of the separation $|i-j|$ between $e_i$ and $\bar{e}_j$, indicating that the $e$ anyons are deconfined.

Since the open loop connecting vertices $i$ and $j$ can be obtained by cutting the closed loops in the original loop configurations $|c_i\rangle$, the confinement of open loops is directly linked to the confinement of closed loops. Under the general $\beta_z$ deformation introduced in Eq.\,\eqref{eq:z3-z-deformed}, each deformed loop configuration $|\tilde{c}_i\rangle$ is related to the bare configuration $|c\rangle$ through the assignment of a weight as follows:
\begin{align}
|c\rangle \rightarrow |\tilde{c}_i\rangle = \exp \left( - \beta_z l_{c_i} \right) |c_i\rangle,
\label{eq:z3-weight}
\end{align}
where $l_{c_i}$ denotes the total length of loops in the configuration $|c\rangle$.

\begin{figure}[ht]
\centering
\setlength{\abovecaptionskip}{5pt}
\includegraphics[width=0.95\textwidth]{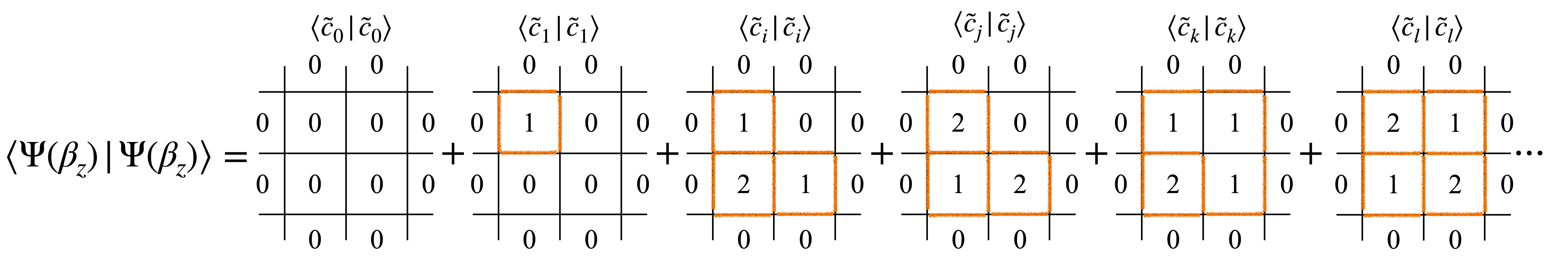}
\caption{Net configurations arising from the overlap $\langle \tilde{\Psi}(\beta_z)|\tilde{\Psi}(\beta_z)\rangle$ under the $\beta_z$ deformation. Each overlap $\langle \tilde{c}_i|\tilde{c}_i\rangle$ is computed from the corresponding loop configuration in Fig.\,\ref{fig:z3-loop}. The directional information of the loops is lost in the overlap, yielding undirected orange nets that may include branching points. The $\mathbb{Z}_3$ numbers on the plaquettes correspond to classical spin variables in the mapping to the $Q=3$ Potts model.
} \label{fig:z3-net-hz}
\end{figure}

To determine whether the closed loops are confined or deconfined under the deformation, one can compute the norm of the wavefunction $\langle \tilde{\Psi}(\beta_z) | \tilde{\Psi}(\beta_z) \rangle$, expressed as:
\begin{revision}
\begin{align}
\langle \tilde{\Psi}(\beta_z) | \tilde{\Psi}(\beta_z) \rangle
= \sum_c \langle \tilde{c} | \tilde{c} \rangle
= \sum_N \chi_N(3) \exp\left(-2\beta_z l_N\right).
\label{eq:z3-hz-norm}
\end{align}
\end{revision}
Here, $\langle \tilde{c} | \tilde{c} \rangle$ represents the overlap of a single configuration, which is reformulated in terms of net configurations, as depicted in Fig.\,\ref{fig:z3-net-hz}. Note that, only the overlaps between the same configuration contribute to the summation due to the orthogonality of the configurations. In this overlap, the directional information from loops is discarded, resulting in nets that may include branching. Consequently, the summation over loop configurations translates to a summation over net configurations $N$, where $l_N$ is the total length of the net, and $\chi_N(3)$ represents the number of loop configurations that correspond to the same net. This mapping is intrinsically connected to the chromatic polynomial $\chi_N(Q)$, which counts the number of ways to color the net $N$ using $Q$ distinct colors. For instance, Fig.\,\ref{fig:z3-net-hz} demonstrates how the overlap of different loop configurations can yield the same net, necessitating a combinatorial enumeration of all loop configurations contributing to that net.

Determining whether the loops in the original loop configurations are confined naturally translates to examining the confinement of nets in the net configurations described by Eq.\,\eqref{eq:z3-hz-norm}. From the discussion earlier one can deduce that in the absence of deformation, the $e$-anyon pair creation operator generates open nets of varying lengths within the overlap configuration $\langle \tilde{c} | \tilde{c} \rangle$. In this deconfined phase, the contributions of all net configurations to the overlap are independent of the lengths of the open nets.

When a length-dependent weight is introduced, however, the contribution of each net configuration becomes sensitive to the length of the open nets. Beyond a certain point, the nets undergo a transition to a confined phase, making it progressively harder to separate the $e$-anyons created by the pair creation operator. This transition directly results in the confinement of the $e$-anyons.

\begin{revision}
For a path of length $R=|i-j|$, define the deformed pair state
$|e_i\bar e_j;\beta_z\rangle
=D(\beta_z)\hat O_{\boldsymbol\gamma}|{\rm TC}_3\rangle$, with
$D(\beta_z)=\prod_lD_l(\beta_z)$. We use the normalized pair-state norm
\begin{align}
\mathcal R_e(R)
&=\frac{\langle e_i\bar e_j;\beta_z|e_i\bar e_j;\beta_z\rangle}
{\langle\tilde\Psi(\beta_z)|\tilde\Psi(\beta_z)\rangle}
\nonumber\\
&=\frac{\langle{\rm TC}_3|\hat O_{\boldsymbol\gamma}^\dagger
D(\beta_z)^2\hat O_{\boldsymbol\gamma}|{\rm TC}_3\rangle}
{\langle{\rm TC}_3|D(\beta_z)^2|{\rm TC}_3\rangle}.
\label{eq:fm-type-line-tension}
\end{align}
Equation~\eqref{eq:fm-type-line-tension} is a wavefunction
Fredenhagen--Marcu-type,
gauge-invariant line-tension diagnostic: the bulk wavefunction normalization
is divided out, while the endpoint-separation dependence measures the cost of
an open defect string. We denote it by $\mathcal R_e$, rather than by the
conventional Fredenhagen--Marcu symbol, because the standard lattice-gauge definition uses a
half-loop divided by the square root of a closed-loop expectation value
~\cite{fredenhagen83,fredenhagen86,gregor11,xu25}. The two constructions serve
the same diagnostic purpose here, but their normalization and phase conventions
should not be conflated.
\end{revision}

\subsubsection{Mapping to Classical Potts Model}

The confinement behavior of the nets described in Eq.\,\eqref{eq:z3-hz-norm} can be analyzed by mapping the problem to a classical $Q=3$ Potts model, characterized by the Hamiltonian:
\begin{align}
H_{\rm Potts} = -J \sum_{\langle i,j \rangle} \delta_{s_i, s_j},
\end{align}
where $s_i$ represents a local $\mathbb{Z}_3$ spin variable. The energy of a given configuration is determined by the length of the domain walls separating regions with distinct spin values, which correspond to closed nets. Thus, the net configurations shown in Fig.\,\ref{fig:z3-net-hz} can be interpreted as classical Potts model configurations with excitations, where each local net excitation is weighted by a Boltzmann factor of $\exp(-\beta)$. In this framework, the partition function of the Potts model can be expressed as:
\begin{align}
Z_{\rm Potts} = \sum_N \chi_N(3) \exp\left(-\beta \, l_N\right),
\label{eq:3potts-partition}
\end{align}
where $\beta = J / T$, $N$ denotes all possible closed-net configurations, $l_N$ is the total length of the nets in a given configuration, and $\chi_N(Q)$ represents the chromatic polynomial.

\begin{revision}
Comparing Eqs.~\eqref{eq:z3-hz-norm} and
\eqref{eq:3potts-partition} gives the parameter dictionary
\begin{align}
\beta=2\beta_z.
\label{eq:potts-dictionary}
\end{align}
The net configurations shown in Fig.~\ref{fig:z3-net-hz} can therefore be
interpreted as configurations of the classical $Q=3$ Potts model, with the
$\mathbb Z_3$ plaquette labels playing the role of classical spins.
\end{revision}

\begin{figure}[ht]
\centering
\setlength{\abovecaptionskip}{5pt}
\includegraphics[width=0.75\textwidth]{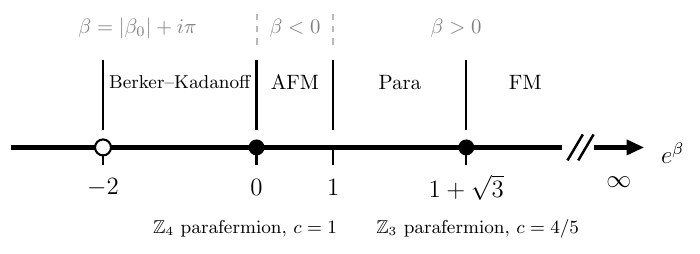}
\caption{{Phase diagram of the $Q=3$ Potts model as a function of
$e^{\beta}$. The ferromagnetic (FM) transition at
$e^{\beta_c}=1+\sqrt3$ is governed by the $\mathbb Z_3$ parafermion conformal
field theory (CFT) with $c=4/5$. The real-temperature antiferromagnetic (AFM)
regime is
$0<e^{\beta}<1$, and its zero-temperature endpoint at $e^{\beta}=0$ is an
isolated $\mathbb Z_4$ parafermion critical point with $c=1$. The inaccessible
imaginary-temperature Berker--Kadanoff phase occupies
$-2<e^{\beta}<0$; $e^{\beta}=-2$ is its opposite endpoint.}}
\label{fig:potts-phase-diagram}
\end{figure}

The \(Q=3\) Potts model exhibits a global \(\mathbb{Z}_3\) symmetry defined by the operation:
\begin{align}
U_1 = \prod_i X_i^s,
\end{align}
where the local operator \(X_i^s\) cyclically rotates the classical spin \(s_i\) through the states \(0 \to 1 \to 2 \to 0\). The phases of the Potts model are characterized by whether, and how, the global symmetry \(U_1\) is spontaneously broken. The overall phase diagram of the \(Q=3\) Potts model, parameterized by \(\exp(\beta)\), is illustrated in Fig.\,\ref{fig:potts-phase-diagram}~\cite{Jacobsen06}. Interestingly, this phase diagram includes the imaginary temperature regime where \(\exp(\beta) < 0\). The real temperature regime is further subdivided into two distinct regions: the ferromagnetic (FM) regime with \(J > 0\) and \(\exp(\beta) \geq 1\), and the antiferromagnetic (AFM) regime with \(J < 0\) and \blue{\(0 < \exp(\beta) < 1\)}.

% This critical point is characterized by a $\mathbb{Z}_3$ conformal field theory (CFT) with central charge $c = \frac{4}{5}$. For $\exp\left(\beta\right) < \exp\left(\beta_c\right)$, the system is symmetric under \(U_1\) and the classical spins are in a disordered paramagnetic phase, resulting in the proliferation of nets that manifest as domain walls. This disordered phase includes the toric code point at $\beta = 0$, which corresponds to the classical system in the high-temperature limit ($T \rightarrow \infty$). In this regime, the $e$ anyons remain deconfined.

In the FM regime, the \(Q=3\) Potts model exhibits a phase transition at:
\begin{align}
\exp\left(\beta_c \right) = 1+\sqrt{3}.
\end{align}
This critical point is governed by a \(\mathbb{Z}_3\) conformal field theory (CFT) with a central charge of \(c = \frac{4}{5}\).
\begin{revision}
Using Eq.~\eqref{eq:potts-dictionary}, its location in the wavefunction
deformation is
\begin{align}
e^{2\beta_{z,c}}=1+\sqrt3,
\qquad
e^{\beta_{z,c}}=\sqrt{1+\sqrt3},
\qquad
t_{z,c}=\tanh(\beta_{z,c}/2)\simeq0.2461.
\label{eq:potts-critical-deformation}
\end{align}
For temperatures above the classical critical value
($e^{\beta}<e^{\beta_c}$, equivalently
$e^{2\beta_z}<1+\sqrt3$), the system remains symmetric under $U_1$, and the
classical spins reside in a disordered paramagnetic phase. In this phase, nets
representing domain walls proliferate throughout the ensemble. This disordered
phase includes the toric-code point $\beta=\beta_z=0$, corresponding to the
high-temperature limit $T\to\infty$.
\end{revision}

As \(\beta\) increases from \(0\), corresponding to a decrease in the classical temperature \(T\), the ensemble undergoes spontaneous symmetry breaking, and the classical spins gradually begin to align. This process continues until the system reaches the critical point at \(\beta = \beta_c\). At this critical point, the system transitions into an ordered phase. In the ordered phase, where \(\exp\left(\beta\right) > \exp\left(\beta_c\right)\), the classical spins become aligned, resulting in confined nets and, consequently, the confinement of \(e\) anyons.

\begin{revision}
In the real-temperature AFM regime $0<e^{\beta}<1$, the classical system
remains disordered and the $e$ anyons remain deconfined; this extended regime
is not itself a critical phase boundary. Its endpoint $e^{\beta}=0$ is the
$T=0$ AFM $Q=3$ Potts critical point, described by the $\mathbb Z_4$
parafermion CFT with $c=1$~\cite{Jacobsen06-1}. Crossing formally to
$e^{\beta}<0$ enters the Berker--Kadanoff phase, but that
imaginary-temperature interval is inaccessible from the real wavefunction
deformation.
\end{revision}

\subsubsection{Emergent $U(1)$ 1-Form Symmetry and Hilbert Space Fragmentation}
\label{sec:emerget-u1}
\blue{At the AFM critical endpoint $\beta_z\to-\infty$ (equivalently
$t_z=-1$ and $e^{\beta}=0$),} the loop weight dominates in the deformed
wavefunction $|\tilde{\Psi}(\beta_z \to -\infty)\rangle$, and configurations containing at least one unoccupied link ($|0\rangle$) are eliminated, leaving only fully packed configurations. These remaining configurations map to the 6-vertex model by associating $|1\rangle$ and $|2\rangle$ states with \blue{right (down) and left (up) arrows}, respectively. At this point,  all vertices in every configuration satisfy the ``two-in-two-out" condition, making it equivalent to the square-ice model.

Configurations with plaquettes featuring clockwise or counterclockwise arrow arrangements are flippable, meaning they can transition to another configuration by flipping the direction of arrows in that plaquette using $b_p$ or $b_p^2$. Conversely, plaquettes where the arrows fail to form a closed cycle are unflippable.

\begin{figure}[ht]
\centering
\setlength{\abovecaptionskip}{5pt}
\includegraphics[width=0.6\textwidth]{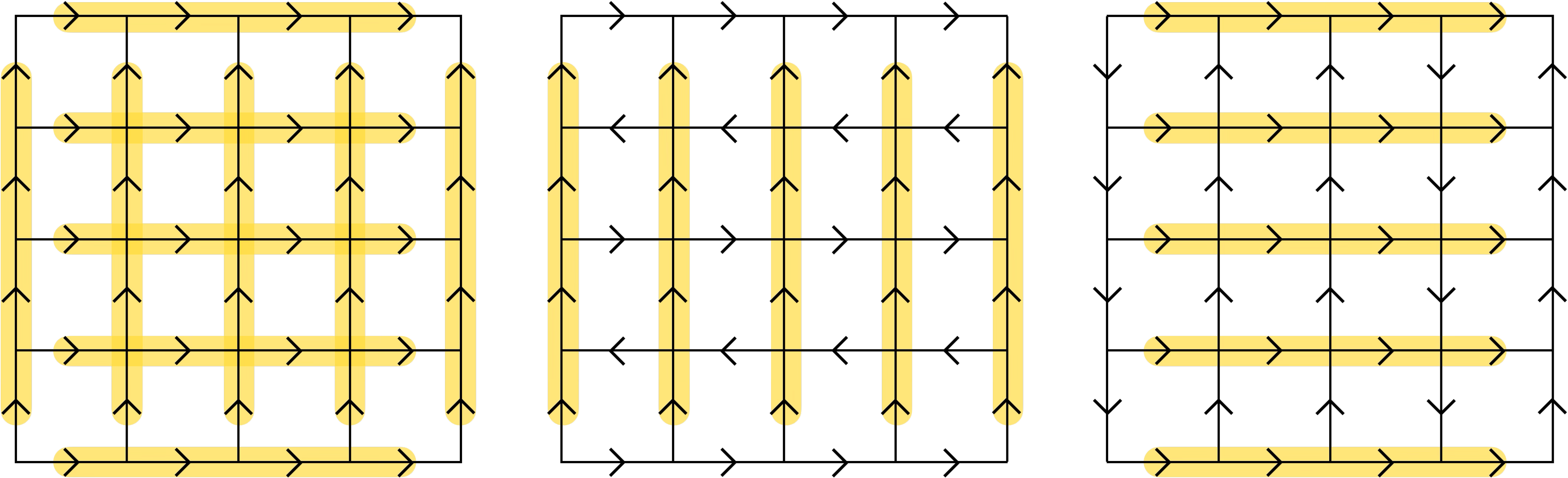}
\caption{{Examples of scar-state configurations at the square-ice
point $\beta_z\to-\infty$ ($t_z=-1$), where all vertices satisfy the
two-in--two-out condition. Yellow shading highlights links that are uniformly
aligned in one direction. When all links along either the vertical or the
horizontal direction are aligned, every plaquette is unflippable under $b_p$
and $b_p^2$, producing configurations disconnected from the rest of the
Hilbert space.}}
\label{fig:2in2out-scar}
\end{figure}

Recent studies\,\cite{stahl24scipost} show that the ensemble of 6-vertex model possesses an emergent $U(1)$ 1-form symmetry. At $\beta_z = -\infty$, the deformed wavefunction acquires this additional symmetry alongside the original 1-form symmetry of the parent Hamiltonian. In the PEPS representation of the deformed wavefunction, this emergent $U(1)$ 1-form symmetry manifests as an injective symmetry of the local tensors. Detailed tensor network analysis is presented in Sec.\,\ref{sec:tn}.

The emergent $U(1)$ 1-form symmetry charge is defined by:
\begin{align}
\hat{N}_{\cal C} = \sum_{l_{\rm odd} \in {\cal C}} \hat{\cal{P}}_l^1 - \sum_{l_{\rm even} \in {\cal C}} \hat{\cal{P}}_l^2,
\label{eq:u1-charge}
\end{align}
\begin{revision}
Here $\hat{\mathcal P}_l^a=|a\rangle_l\langle a|_l$ for $a=1,2$.
The non-contractible dual-lattice cut $\mathcal C$ alternately crosses the two
link orientations, denoted by the odd and even link families. With the arrow
convention of Fig.~\ref{fig:z3-loop}, the relative minus sign converts the two
local orientation conventions into flux measured in one common direction.
Thus $\hat N_{\mathcal C}$ is the net signed arrow flux through the cut. A
local plaquette move $b_p$ or $b_p^2$ changes the arrows at the two
intersections with any closed cut by equal and opposite amounts, so
\begin{align}
[\hat N_{\mathcal C},b_p]=[\hat N_{\mathcal C},b_p^2]=0.
\end{align}
The charge consequently depends only on the homotopy class of $\mathcal C$,
which is the defining one-form conservation law. Its extremal values are
$\pm L/2$, in agreement with the uniformly aligned, unflippable scar
configurations in Fig.~\ref{fig:2in2out-scar}.
\end{revision}

This emergent symmetry fragments the Hilbert space, giving rise to ``scar'' state configurations whose number grows exponentially with the system size $L$ under periodic boundary conditions. These configurations have no flippable plaquettes and exhibit the maximum absolute value of the symmetry charge, $\hat{N}_{\cal C} = \pm L/2$. Examples of such configurations are illustrated in Fig.\,\ref{fig:2in2out-scar}. Additional configurations can be constructed as follows. In the states shown in Fig.\,\ref{fig:2in2out-scar}, either all vertical links or all horizontal links are aligned in the same direction, indicated by the yellow shading. When all links in either the vertical or horizontal direction are uniformly aligned, all plaquettes become unflippable. For example, if all vertical links are aligned, the $L$ rows of links gain freedom to choose their directions independently. This results in $2^{L+1}$ configurations. Rotating the lattice by $90^\circ$ produces $2^{L+1}$ additional configurations, all distinct from the original set. After accounting for overlaps, the total number of scar states is $2^{L+2} - 4$.

\subsection{Phase Diagram of $|\tilde{\Psi}(\beta_x)\rangle$}
\subsubsection{Loop and Net Description}
The dual wavefunction $|\tilde{\Psi}(\beta_x)\rangle$ can be analyzed in a similar fashion. In the limit $\beta_x \rightarrow 0$, the bare TC wavefunction emerges, where the local basis states $|0\rangle$, $|1\rangle$, and $|2\rangle$ remain orthonormal. As $\beta_x$ increases, the orthonormality of the local basis is lost, as described by Eq.\,\eqref{eq:beta-x-local-overlap}. In the opposite limit, $\beta_x \rightarrow \infty$, the three basis states align in the same direction, causing the loop configurations $\{ |\tilde{c}_i\rangle\}$ to become indistinguishable. In this regime, $e$ anyons are condensed, as creating an open loop in any configuration results in a state identical to the original.

More precisely, consider a loop configuration $|\tilde{c}\rangle$ and the configuration $|\tilde{c}'\rangle$ obtained by creating an $e$-anyon pair. The overlap $\langle \tilde{c}| \tilde{c}'\rangle$ becomes non-zero when $\beta_x$ is finite. Along the path connecting the sites where anyons are created, $i$ and $j$, the links correspond to the overlap of two distinct local states, $\langle \tilde{a}|\tilde{b}\rangle \equiv \mu$, where $a \neq b$. Without deformation, the overlap $\langle \tilde{c}| \tilde{c}'\rangle$ vanishes because the discrepancy line contributes $\mu = 0$. However, under $\beta_x$ deformation, this contribution becomes non-zero, resulting in $\langle \tilde{c}| \tilde{c}'\rangle \neq 0$. 

As a result, in the limit $\beta_x \rightarrow \infty$ (hence $\mu\rightarrow 1$), $\langle \tilde{\Psi}(\beta_x) | e_i \bar{e}_j\rangle \neq 0$, indicating that the $e$ anyons are condensed. This occurs because the discrepancy lines become deconfined. The condensation of $e$ anyons is directly tied to the proliferation of these discrepancy lines. At the phase transition point, located somewhere in the intermediate regime where the $e$ anyons begin to condense, the discrepancy lines associated with $\mu = \langle \tilde{a}|\tilde{b}\rangle$ must also start to deconfine.

\begin{figure}[ht]
\centering
\setlength{\abovecaptionskip}{5pt}
\includegraphics[width=1\textwidth]{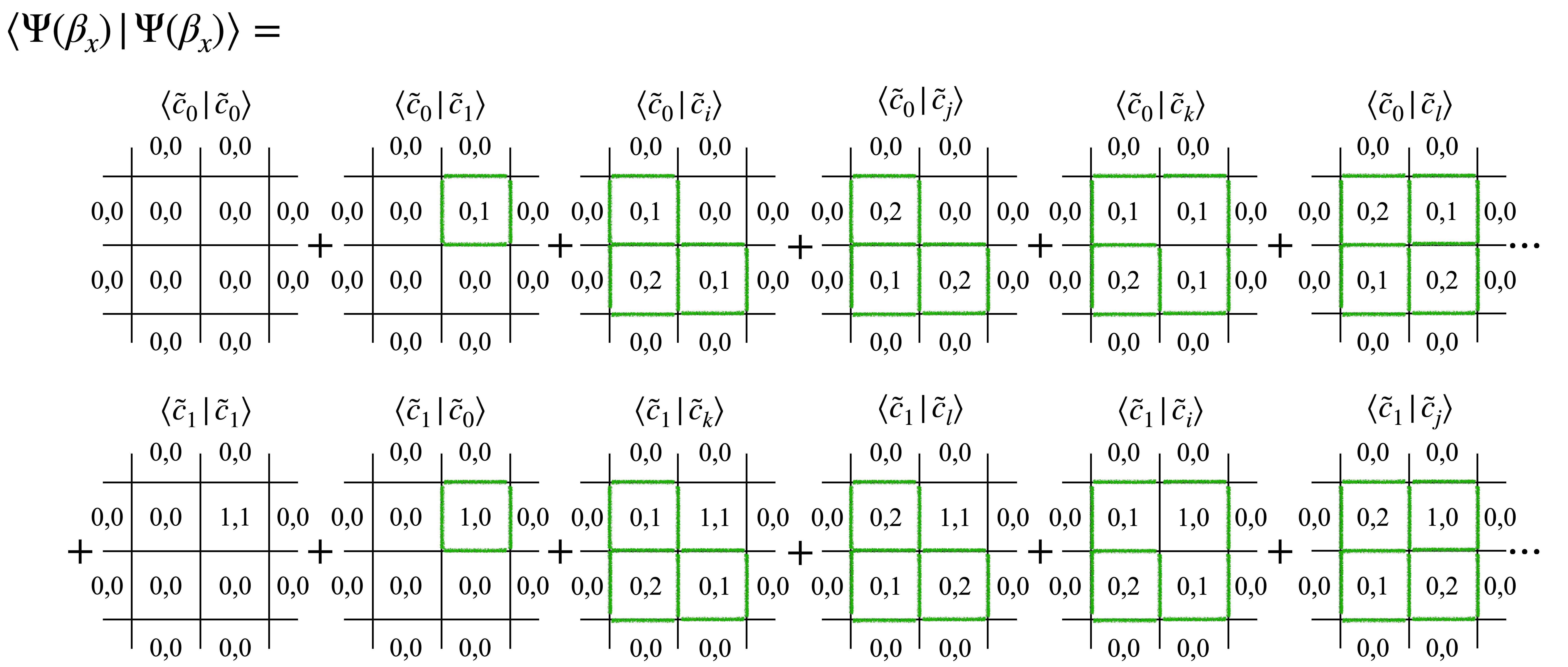}
\caption{Net configurations arising from the overlap $\langle \tilde{\Psi}(\beta_x)|\tilde{\Psi}(\beta_x)\rangle$ under the $\beta_x$ deformation. The first row shows overlaps of all loop configurations with the vacuum state $|c_0\rangle$; the second row shows overlaps with $|c_1\rangle$. Each plaquette carries a pair of $\mathbb{Z}_3$ numbers $(s_i, \sigma_i)$ from the ket and bra configurations. Green lines indicate discrepancy lines where the local overlap $\mu = \langle \tilde{a}|\tilde{b}\rangle$ ($a \neq b$) contributes. The two rows yield the same set of net configurations.
} \label{fig:z3-net-hx}
\end{figure}

Discrepancy lines also appear in the overlap of two distinct closed loop configurations, $\langle \tilde{c}_i | \tilde{c}_j\rangle$, which becomes non-zero under the deformation $\beta_x$. Consequently, the overlaps of different configurations contribute to the norm of the deformed wavefunction:
\begin{align}
\langle \tilde{\Psi}(\beta_x) | \tilde{\Psi}(\beta_x)\rangle  = \sum_{c_i, c_j} \langle \tilde{c}_i | \tilde{c}_j\rangle.
\label{eq:x-deform-norm-1}
\end{align}
As shown in Fig.\,\ref{fig:z3-net-hx}, these discrepancy lines manifest as closed nets. Each plaquette now contains two numbers: one from the ket configuration and one from the bra configuration of the closed loops. The nets appear on the links that exhibit discrepancies in the local state overlaps, $\mu = \langle \tilde{a}|\tilde{b}\rangle$. The first and second rows in Fig.\,\ref{fig:z3-net-hx} depict the overlap configurations of all closed loop states with the states $|c_0\rangle$ and $|c_1\rangle$, respectively. Notably, the first and second rows yield the same set of net configurations. 

In this way, the set of overlaps between any closed loop configuration $|c_i\rangle$ and all other configurations is equivalent to the set of overlaps between the vacuum state $|c_0\rangle$ and all configurations. Consequently, Eq.\,\eqref{eq:x-deform-norm-1} can be expressed as:
\begin{align}
\langle \tilde{\Psi}(\beta_x) | \tilde{\Psi}(\beta_x)\rangle  =  N_N \sum_{N} \chi_N(3) \mu^{l_N}.
\label{eq:x-deform-norm-2}
\end{align}
Here, $N_N$ denotes the total number of net configurations, $l_N$ is the total length of the nets in a given configuration, and $\chi_N(Q)$ is the chromatic polynomial.

\subsubsection{Mapping to Classical Potts Model}

The analysis of whether the nets in Eq.\,\eqref{eq:x-deform-norm-2} are confined or not also can be mapped to the study of classical $Q=3$ Potts model as well. The norm of wavefunction can be mapped to classical partition function in Eq.\,\eqref{eq:3potts-partition} by matching $\mu = \exp(-\beta)$. 

Notably, by labeling the two digits on each plaquette in Fig.\,\ref{fig:z3-net-hx} as \(s_i\) and \(\sigma_i\), the nets emerge as domain walls separating plaquettes where the difference between the two \(\mathbb{Z}_3\) numbers, \(s_i - \sigma_i\), is nonzero. Thus, each net configuration can be interpreted as a classical configuration of the \(Q=3\) Potts model by treating the difference of the two \(\mathbb{Z}_3\) numbers, \(S_i \equiv s_i - \sigma_i\), as a classical spin. The corresponding classical model is described by:
\begin{align}
H_{\rm Potts}' = -J' \sum_{\langle i,j \rangle} \delta_{S_i, S_j},
\end{align}
This model exhibits a global \(\mathbb{Z}_3\) symmetry defined by the operation:
\begin{align}
U_2 = \prod_i X_i^s X_i^\sigma,
\end{align}
where the local operators \(X_i^s\) and \(X_i^\sigma\) cyclically rotate the classical spins \(s_i\) and \(\sigma_i\) through the states \(0 \to 1 \to 2 \to 0\), respectively. The phases of the Potts model are characterized by whether, and how, the global symmetry \(U_2\) is spontaneously broken.

The variational parameter \(\mu\) spans the range \(-\frac{1}{2}\) to \(1\). For nonzero values of \(\mu \geq 0\), the corresponding classical system resides in the FM regime, where a phase transition occurs at:
\begin{align}
\mu_c = \frac{1}{1+\sqrt{3}}.
\end{align}
In the range \(0 \leq \mu < \mu_c\), below this critical point, the classical system shows spontaneous symmetry breaking of \(U_2\) and remains in an ordered phase, with the discrepancy lines suppressed. Consequently, the \(e\) anyons in the deformed toric code phase remain uncondensed within this parameter regime. Conversely, for \(\mu_c < \mu \leq 1\), above the critical point, the classical system transitions to a disordered phase characterized by the proliferation of discrepancy lines. This transition results in the condensation of \(e\) anyons in the deformed toric code phase within this parameter range.

For negative values of \(\mu\), the classical system enters an imaginary temperature regime. In this case, the system exhibits AFM ordering, except at the second critical point \(\mu_{c'} = -\frac{1}{2}\). Similar to the FM phase, the discrepancy lines are suppressed in the AFM phase, as they emerge as domain walls between plaquettes with differing $S_i$ values. As a result, \(e\) anyons do not condense in this phase. Therefore, the phases of the deformed TC corresponding to the FM and AFM regimes, \(\mu_{c'} < \mu < \mu_c\), can be regarded as belonging to the same phase.

\subsubsection{Square Ice in Dual System}

In the limit \(\beta_x \to -\infty\) or \(\mu = -\frac{1}{2}\), the system reaches a critical state described by \(\mathbb{Z}_4\) parafermion conformal field theory (CFT). At this point, the three local vectors, \(|\tilde{0}_l\rangle\), \(|\tilde{1}_l\rangle\), and \(|\tilde{2}_l\rangle\), lie in a two-dimensional subspace and are symmetrically related by rotations of \(\frac{2\pi}{3}\). This relationship is captured by the condition \(|\tilde{0}_l\rangle + |\tilde{1}_l\rangle + |\tilde{2}_l\rangle = 0\). Consequently, in this limit, the local Hilbert space required to describe the deformed wavefunction reduces from dimension 3 to dimension 2.

Under the \({\cal D}_{\rm e-m}\) duality mapping, the \(\beta_x\)-deformed wavefunction is transformed into a \(\beta_z\)-deformed wavefunction. At the dual point where \(\beta_z \to -\infty\), the wavefunction corresponds to a fully packed loop state, with the local Hilbert space spanned by two orthonormal basis states. Within this dual framework, the three local vectors transform as follows:
\begin{align}
|\tilde{0}_l\rangle &\to |\tilde{\omega}^0_l\rangle = \frac{1}{\sqrt{3}}\left(|\tilde{0}_l\rangle + |\tilde{1}_l\rangle + |\tilde{2}_l\rangle\right) = 0, \notag \\
|\tilde{1}_l\rangle &\to |\tilde{\omega}^1_l\rangle = \frac{1}{\sqrt{3}}\left(|\tilde{0}_l\rangle + \omega |\tilde{1}_l\rangle + \omega^2|\tilde{2}_l\rangle\right), \notag \\
|\tilde{2}_l\rangle &\to |\tilde{\omega}^2_l\rangle = \frac{1}{\sqrt{3}}\left(|\tilde{0}_l\rangle + \omega^2|\tilde{1}_l\rangle + \omega|\tilde{2}_l\rangle\right),
\end{align}
where \(|\tilde{\omega}^1_l\rangle\) and \(|\tilde{\omega}^2_l\rangle\) are orthonormal. Using this new basis, the original three vectors before the dual mapping can be expressed as:
\begin{align}
|\tilde{0}_l\rangle &= \frac{\sqrt{2}}{2} \left(|\tilde{\omega}_l^1\rangle + |\tilde{\omega}_l^2\rangle\right), \notag \\
|\tilde{1}_l\rangle &= \frac{\sqrt{2}}{2} \left(\omega^2 |\tilde{\omega}_l^1\rangle + \omega|\tilde{\omega}_l^2\rangle\right), \notag \\
|\tilde{2}_l\rangle &= \frac{\sqrt{2}}{2} \left(\omega|\tilde{\omega}_l^1\rangle + \omega^2|\tilde{\omega}_l^2\rangle\right).
\end{align}

Since the dual local basis \(|\tilde{\omega}^1\rangle\) and \(|\tilde{\omega}^2\rangle\) span the two-dimensional local Hilbert space formed by the three deformed vectors \(|\tilde{0}\rangle\), \(|\tilde{1}\rangle\), and \(|\tilde{2}\rangle\), the set of configurations of the dual lattice, \(\{|\tilde{d}_i\rangle\}\), where every link is assigned the local basis \(|\tilde{\omega}^1\rangle\) or \(|\tilde{\omega}^2\rangle\), constitutes a complete set encompassing all configurations in \(|\tilde{\Psi}(\beta_x \to -\infty)\rangle\). One can also interpret the dual configuration \(|\tilde{d}_i\rangle\) as a loop configuration by treating \(|\tilde{\omega}^1\rangle\) as representing an upward (or leftward) loop and \(|\tilde{\omega}^2\rangle\) as representing a downward (or rightward) loop.

In the dual basis, the deformed wavefunction can be decomposed as
\begin{align}
|\tilde{\Psi}(\beta_x \to -\infty)\rangle = 
% \sum_i |\tilde{d}_i\rangle \langle \tilde{d}_i|\tilde{\Psi}(\beta_x \to -\infty)\rangle = 
\sum_{j} \left( \sum_i \langle \tilde{d}_i |\tilde{c}_j\rangle \right) |\tilde{d}_i\rangle.
\end{align}
Here, the overlap of configurations \(\sum_j \langle \tilde{d}_i |\tilde{c}_j\rangle\) becomes zero when the dual loop configuration contains open loops. However, the overlaps with dual configurations composed entirely of closed loops are equal to each other.

To demonstrate this, consider the overlaps between the local vectors, which are given by:
\begin{align}
\langle \tilde{\omega}^1| \tilde{0}\rangle &= \langle \tilde{\omega}^2| \tilde{0}\rangle = 1, \notag \\
\langle \tilde{\omega}^1| \tilde{1}\rangle &= \langle \tilde{\omega}^2| \tilde{2}\rangle = \omega^2, && \text{(clockwise)} \notag \\
\langle \tilde{\omega}^2| \tilde{1}\rangle &= \langle \tilde{\omega}^1| \tilde{2}\rangle = \omega. && \text{(counterclockwise)}
\end{align}
As indicated above, the overlaps between the dual loop states and the original loop states become \(\omega^2\) when the direction of the dual loop corresponds to a clockwise \(90^\circ\) rotation of the original loop's direction. Similarly, the overlaps become \(\omega\) when the dual loop's direction corresponds to a counterclockwise \(90^\circ\) rotation of the original loop's direction. The overlap between the dual loop state and the unoccupied original link state is \(1\).

Next, consider a loop in a certain configuration \(|\tilde{d}_i\rangle\). In its overlap with the deformed wavefunction, we have
\begin{align}
\langle \tilde{d}_i|\tilde{\Psi}(\beta_x \to -\infty)\rangle = \sum_j \langle \tilde{d}_i | \tilde{c}_j\rangle,
\end{align}
where the dual loop in \(|\tilde{d}_i\rangle\) interacts with the original loops in \(|\tilde{c}_j\rangle\). Focusing on the region supported by the dual loop under consideration, note that since the original loops are closed by construction, any dual loop in closed form necessarily crosses each original closed loop twice\blue{---}once in a clockwise direction and once in a counterclockwise direction. Consequently, the contribution of the dual loop to each \(\langle \tilde{d}_i | \tilde{c}_j\rangle\) is \(1\).

On the other hand, when the dual loop under consideration has open ends, each end of the dual loop can be located within the interior of an original closed loop. This results in a single crossing, either clockwise or counterclockwise, contributing either \(\omega^2\) or \(\omega\) to the overlap configuration. Since there exist an equal number of closed loop configurations where the closed loop enclosing the dual line ends is either absent or circulates in opposite directions, the contributions of these loops cancel out. Specifically, the overlaps with the dual open line yield \(1\), \(\omega^2\), or \(\omega\), and the overall summation \(\sum_j \langle \tilde{d}_i | \tilde{c}_j \rangle\) vanishes because \(1 + \omega + \omega^2 = 0\). In other words, dual configurations containing an open dual line have zero overlap with the deformed wavefunction.

As a result, the deformed wavefunction in the limit of \(\beta_x\rightarrow - \infty\) limit can be expressed by equal superposition of (fully packed) closed loop configuration as follows:
\begin{align}
|\tilde{\Psi}(\beta_x \rightarrow -\infty)\rangle \propto \sum_i |\tilde{d}_i\rangle,
\end{align}
which is equivalent to the two-in-two-out configurations in Sec.\,\ref{sec:emerget-u1}.

\section{Tensor Network Analysis}
\label{sec:tn}

In the previous section, we introduced two types of deformations to the \(\mathbb{Z}_3\) toric code (TC), labeled as \(\beta_x\) and \(\beta_z\) deformations. Utilizing the loop and net configuration framework, we analyzed the phase diagrams for each deformation. In each case, we identified phases where the \(e\) anyons are either condensed or confined, respectively. Additionally, in both cases, we observed a critical point exhibiting characteristics of the square ice model with an emergent \(U(1)\) symmetry. Naturally, this raises questions about how the \(e\)-condensed and \(e\)-confined phases are connected, and what other phases might exist near the square ice critical point.

To address these questions, we introduce in this section a generalized deformed \(\mathbb{Z}_3\) TC wavefunction characterized by two deformation parameters, \(\beta_x\) and \(\beta_z\), and analyze it using the tensor network (TN) methodology. By expressing the norm of the wavefunction as a product involving a one-dimensional transfer matrix, we reveal that one of the holonomy operators acts as a global symmetry of the 1D system described by this transfer matrix. The breaking of this global symmetry is directly linked to the condensation or confinement of the \(e\) anyons. Furthermore, at the square ice critical point, an emergent \(U(1)\) symmetry appears as an additional global symmetry.

\begin{revision}
This section first introduces the two-parameter deformed $\mathbb Z_3$ TC
wavefunction and then maps its norm to the three-state Ashkin--Teller-like
(AT$_3$) construction, in which the two four-spin constraints that coincide
for $\mathbb Z_2$ become independent. We finally use the variational uniform
matrix-product-state (VUMPS) method to analyze the corresponding transfer
matrix in the thermodynamic limit.
\end{revision}

\subsection{Generalized Deformed \(\mathbb{Z}_3\) TC}

We start by defining the generalized deformed \(\mathbb{Z}_3\) TC wavefunction as follows:
\begin{align}
|\tilde{\Psi}(\beta_x, \beta_z)\rangle = \prod_l D_l(\beta_x, \beta_z) |{\rm TC}_3\rangle,
\end{align}
where the deformation operator is given by
\begin{align}
D_l(\beta_x, \beta_z) \equiv  
I_l + \tanh\left(\frac{\beta_x}{2}\right) \Gamma_l^x + \tanh\left(\frac{\beta_z}{2}\right) \Gamma_l^z.
\end{align}

This deformation transforms the local basis states \(|0\rangle\), \(|1\rangle\), and \(|2\rangle\) as:
\begin{align}
|0\rangle \xrightarrow[]{D(\beta_x,\beta_z)} &
|\tilde{0}\rangle \propto 
\left(1 - \frac{1}{3}t_x + t_z\right) |0\rangle 
+ \frac{2}{3} t_x \left(|1\rangle + |2\rangle\right), \nn 
|1\rangle \xrightarrow[]{D(\beta_x,\beta_z)} &
|\tilde{1}\rangle \propto 
\left(1 - \frac{1}{3}t_x - t_z \right)|1\rangle 
+ \frac{2}{3} t_x\left(|0\rangle + |2\rangle\right), \nn 
|2\rangle \xrightarrow[]{D(\beta_x,\beta_z)} &
|\tilde{2}\rangle \propto 
\left(1 - \frac{1}{3}t_x - t_z\right) |2\rangle 
+ \frac{2}{3}t_x \left(|0\rangle + |1\rangle\right).
\end{align}
Here, we introduce two variational parameters, \((t_x, t_z) \equiv \left(\tanh\frac{\beta_x}{2}, \tanh\frac{\beta_z}{2}\right)\), to simplify the expressions.

Normalizing $\langle \tilde{0} |\tilde{0}\rangle  = 1$, the overlaps of local basis are given by
\begin{align}
\langle \tilde{1} | \tilde{1} \rangle  
= &
\langle \tilde{2} | \tilde{2} \rangle = 
\frac{8 t_x^2   + \left( 3 t_z  + t_x - 3 \right)^2}
{8 t_x^2+ \left(3  t_z - t_x + 3 \right)^2} \equiv \nu,
\nn 
\langle \tilde{0} | \tilde{1} \rangle = & \langle \tilde{0} | \tilde{2} \rangle 
= 
\frac{12 t_x}
{8 t_x^2 + \left( 3 t_z - t_x  + 3 \right)^2} \equiv \mu, 
\nn 
\langle \tilde{1}| \tilde{2} \rangle = &  
\frac{ 12 t_x \left( 1  - t_z \right)}
 {8 t_x^2 + \left( 3 t_z - t_x  + 3 \right)^2} \equiv \rho.
\label{eq:z3-loc-overlap}
\end{align}

Previously, we have shown that $\Gamma^x$ and $\Gamma^z$ transform as $\Gamma^x \leftrightarrow \Gamma^z$ under the duality ${\cal D}_{\rm e-m}$ defined in Eq.\,(\ref{eq:toric-code-dual}). As a result, the deformed $\mathbb{Z}_3$ toric code exhibits a $t_x \leftrightarrow t_z$ duality, such that $|\tilde{\Psi}(t_x, t_z)\rangle$ maps to $|\tilde{\Psi}(t_z, t_x)\rangle$. Consequently, the phase diagram of the deformed wavefunction is symmetric about the $t_z = t_x$ line, as shown in Fig.\,\ref{fig:z3tc-phase-diagram}.

\begin{revision}
Unlike the $\mathbb Z_2$ case reviewed in Appendix~\ref{sec:z2-review}, the
deformed $\mathbb Z_3$ wavefunction does not possess a sign-change duality. For
qubits, $\{X,Z\}=0$ relates $|\psi(h_x,h_z)\rangle$ and
$|\psi(-h_x,h_z)\rangle$ unitarily and folds the phase diagram about its axes.
The qutrit relation $ZX=\omega XZ$ provides no analogous folding. Within the
accessible wavefunction parameters, the concrete consequence is the survival
of the isolated AFM endpoints at $t_x=-1$ and $t_z=-1$.
\end{revision}

\subsection{Mapping to a classical Ashkin--Teller-like model}
\begin{figure}
    \centering
    \includegraphics[width=0.7\linewidth]{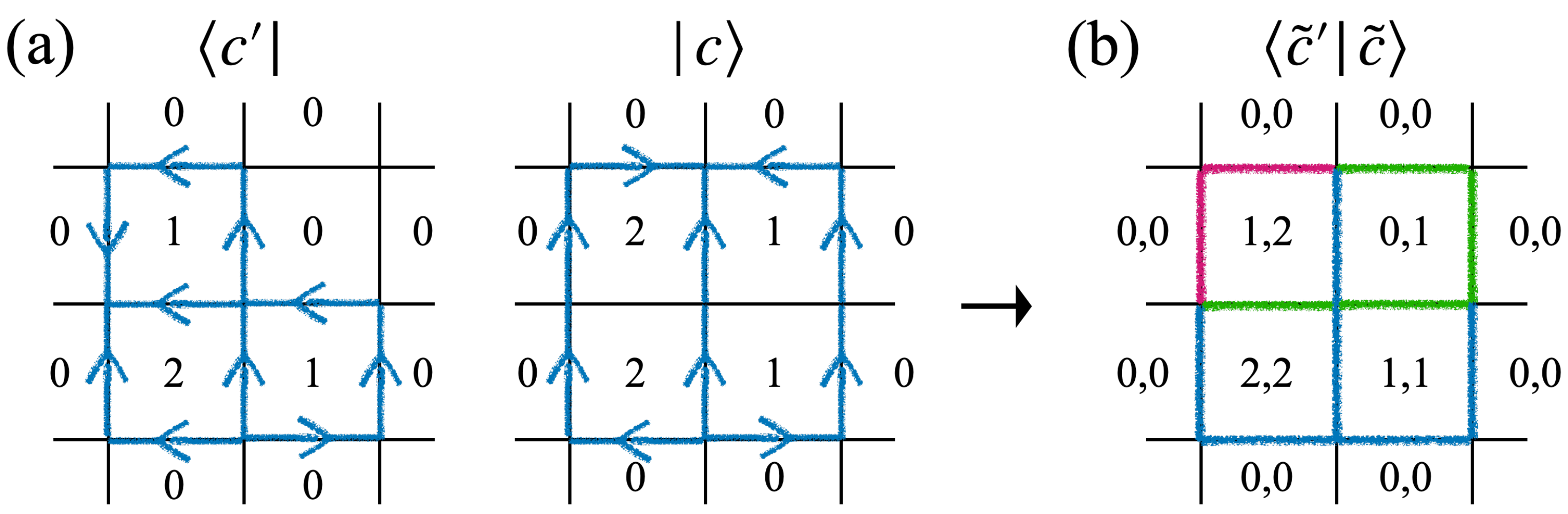}
    \caption{Illustration of the overlap $\langle \tilde{c}'|\tilde{c}\rangle$ between two closed-loop configurations. (a) Two example configurations $\langle c'|$ (left) and $|c\rangle$ (right) on the honeycomb lattice, where the arrows indicate the $\mathbb{Z}_3$ loop orientation and each plaquette is labeled by its $\mathbb{Z}_3$ charge. (b) The resulting overlap $\langle \tilde{c}'|\tilde{c}\rangle$, where each plaquette carries a pair of numbers from the bra and ket configurations. Blue lines mark links where loops from $c'$ and $c$ overlap with parallel alignment, contributing to $l_{c'\parallel c}$. Magenta lines mark links where loops overlap with anti-parallel alignment, contributing to $l_{c'\Updownarrow c}$. Green lines mark links occupied by a loop from only one of the two configurations.}
    \label{fig:at3-overlap}
\end{figure}

The phase diagram of the generalized deformed \(\mathbb{Z}_3\) toric code (TC) can be analyzed by examining the norm of the wavefunction:
\begin{align}
\langle \tilde{\Psi}(t_x, t_z) | \tilde{\Psi}(t_x, t_z) \rangle
= \sum_{c_i, c_j} \langle \tilde{c}_i | \tilde{c}_j \rangle
= \sum_{c_i, c_j} \left[\nu\right]^{l_{c_i \parallel c_j}} \left[\rho\right]^{l_{c_i \Updownarrow c_j}} \left[\mu\right]^{l_{c_i} + l_{c_j} - l_{c_i \parallel c_j} - l_{c_i \Updownarrow c_j}}.
\label{eq:z3tc-xz-norm}
\end{align}
Here, \(|\tilde{c}_i\rangle = D_l(t_x, t_z) | c_i \rangle\) represents closed-loop configurations with the deformation \(D_l(t_x, t_z)\) applied. The term \(l_{c_i}\) denotes the total loop length in the configuration \(c_i\), \(l_{c_i \parallel c_j}\) represents the length of overlapping loops aligned in the same direction between \(c_i\) and \(c_j\), and \(l_{c_i \Updownarrow c_j}\) represents the length of overlapping loops aligned in opposite directions between \(c_i\) and \(c_j\). The summation over these overlap configurations is depicted in Fig.\,\ref{fig:at3-overlap}. In this figure, blue lines indicate links where loops from \(c_i\) and \(c_j\) overlap with parallel alignment, contributing a total length of \(l_{c_i \parallel c_j}\). Magenta lines indicate links where loops from \(c_i\) and \(c_j\) overlap with anti-parallel alignment, contributing a total length of \(l_{c_i \Updownarrow c_j}\). Green lines represent links occupied by a loop from either \(c_i\) or \(c_j\) but not both, contributing a total length of \(l_{c_i} + l_{c_j} - l_{c_i \parallel c_j} - l_{c_i \Updownarrow c_j}\). Each plaquette contains two numbers: one from the ket configuration and the other from the bra configuration of the closed loop.

From the previous discussion, it follows that the confinement of \(e\) anyons is associated with the suppression of blue lines, which correspond to the local overlap \(\nu = \langle \tilde{1}|\tilde{1}\rangle = \langle \tilde{2}|\tilde{2}\rangle\). In contrast, the condensation of \(e\) anyons is linked to the proliferation of magenta or green lines, which correspond to the local discrepancy overlaps \(\mu = \langle \tilde{0}|\tilde{1}\rangle = \langle \tilde{0}|\tilde{2}\rangle\) or \(\rho = \langle \tilde{1}|\tilde{2}\rangle\).

One can get further insights by mapping the norm of deformed wavefunction in Eq.\,\eqref{eq:z3tc-xz-norm} into the classical partition function of which Hamiltonian is defined as following
\begin{align}
H_{{\rm AT}_3} = 
-J_2 \sum_{\langle i,j\rangle} \left( \delta_{s_i, s_j} + \delta_{\sigma_i, \sigma_j}\right) 
- J_{4,1}\sum_{\langle i, j\rangle } \left(\delta_{s_i - \sigma_i, s_j - \sigma_j } \right)
- J_{4,2}\sum_{\langle i, j\rangle } \left(\delta_{s_i + \sigma_i, \sigma_j+s_j} \right).
\end{align}
This model, referred to as AT\(_3\), serves as the \(\mathbb{Z}_3\) generalization of Ashkin--Teller model. Each site \(i\) is defined by two \(\mathbb{Z}_3\) spins $s_i$ and $\sigma_i$. 

By associating \((s_i, \sigma_i)\) with pairs of \(\mathbb{Z}_3\) numbers, Fig.\,\ref{fig:at3-overlap} also represents classical configurations of the AT\(_3\) model. In this classical model, local excitations depend on the differences between neighboring site variables. These excitations correspond to the three types of colored lines observed in the overlap of loop configurations \(\langle \tilde{c}_i| \tilde{c}_j \rangle\) [Fig.\,\ref{fig:at3-overlap}(b)], with their Boltzmann weights mapping onto the quantum system as follows:
\begin{align}
\text{(Blue line)}& & \exp\left(-2 \beta_2 - \beta_{4,2} \right) & = \nu, \nn
\text{(Magenta line)}& & \exp\left(-2 \beta_2  - \beta_{4,1}\right) & = \mu, \nn
\text{(Green line)}& & \exp\left(-\beta_2 - \beta_{4,1} - \beta_{4,2}\right) & = \rho.
\label{eq:at-z3tc-map}
\end{align}
Here, \(\beta_2 = J_2 / T\), \(\beta_{4,1} = J_{4,1}/T\), and \(\beta_{4,2} = J_{4,2}/T\), where \(T\) represents temperature. Using the mapping defined in Eq.\,\eqref{eq:at-z3tc-map}, the classical partition function can be related to the norm of the deformed \(\mathbb{Z}_3\) TC wavefunction as \(Z_{{\rm AT}_3} = \langle \tilde{\Psi}(\beta_x, \beta_z) | \tilde{\Psi}(\beta_x, \beta_z) \rangle\).

The AT\(_3\) model possesses two evident global symmetries, defined as:
\begin{align}
U_1 = \prod_i X_i^s, 
\quad \text{and} \quad
U_2 = \prod_i X_i^s X_i^\sigma,
\label{eq:at3-global-sym}
\end{align}
where \(X_i^s\) and \(X_i^\sigma\) cyclically rotate the local \(\mathbb{Z}_3\) variables \(s_i\) and \(\sigma_i\), respectively. The order parameters corresponding to the spontaneous breaking of each symmetry are given by:
\begin{align}
\langle \omega^{s_i} \omega^{\sigma_i} \rangle, \quad \text{and} \quad
\langle \omega^{\sigma_i} \rangle.
\end{align}
Here, \(\omega = \exp\left(2\pi i / 3\right)\).

The order parameter \(\langle \omega^{s_i} \omega^{\sigma_i} \rangle\) characterizes the ordering of the two \(\mathbb{Z}_3\) numbers. When these two \(\mathbb{Z}_3\) numbers are ordered, the discrepancy lines represented by \(\mu\) and \(\rho\) are suppressed, indicating that the \(e\) anyons remain uncondensed in this phase. 

Conversely, the order parameter \(\langle \omega^{\sigma_i} \rangle\) reflects the ordering of the \(\mathbb{Z}_3\) number \(\sigma_i\). Due to the symmetry of the model under the exchange of \(s_i\) and \(\sigma_i\), the ordering of \(\sigma_i\) implies the ordering of \(s_i\) as well. Furthermore, as confirmed through numerical analysis and shown in Fig.\,\ref{fig:z3tc-phase-diagram}, the order parameter \(\langle \omega^{\sigma_i} \rangle\) or \(\langle \omega^{s_i} \rangle\) can have a finite value only when \(\langle \omega^{s_i} \omega^{\sigma_i} \rangle \neq 0\). This indicates that the discrepancy lines are suppressed in the net configuration when either \(s_i\) or \(\sigma_i\) is ordered. In this scenario, the configuration contains only the blue nets represented by \(\nu\). When \(\langle \omega^{\sigma_i} \rangle \neq 0\), the blue lines are further suppressed, signifying that the \(e\) anyons are confined.

\subsection{PEPS Representation and Injective Symmetries}
The tensor network framework establishes a connection between the classical order parameters and the condensation or confinement of \(e\) anyons in the deformed \(\mathbb{Z}_3\) TC. We begin by introducing the PEPS representation for the \(\mathbb{Z}_3\) TC:
\begin{align}
    \includegraphics[width=0.3\textwidth]{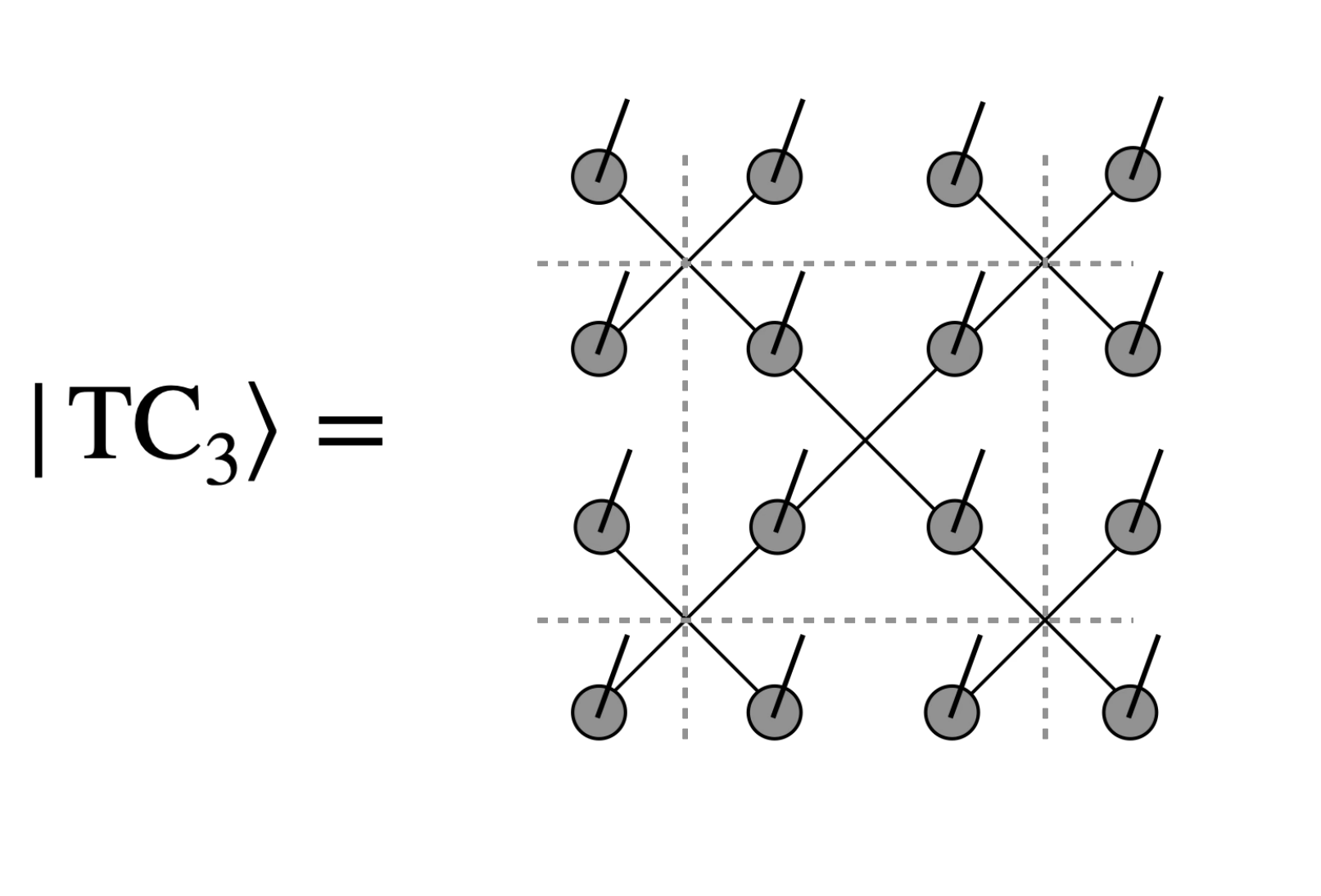}
    \label{eq:peps-z3tc}
\end{align}
In this representation, the solid lines denote the square lattice shown in Fig.\,\ref{fig:tc-model}, which is obtained by rotating the original square lattice \(45\degree\) counterclockwise. The PEPS representation is constructed by contracting the virtual indices of the local PEPS tensor defined as:
\begin{align}
    \includegraphics[width=0.3\textwidth]{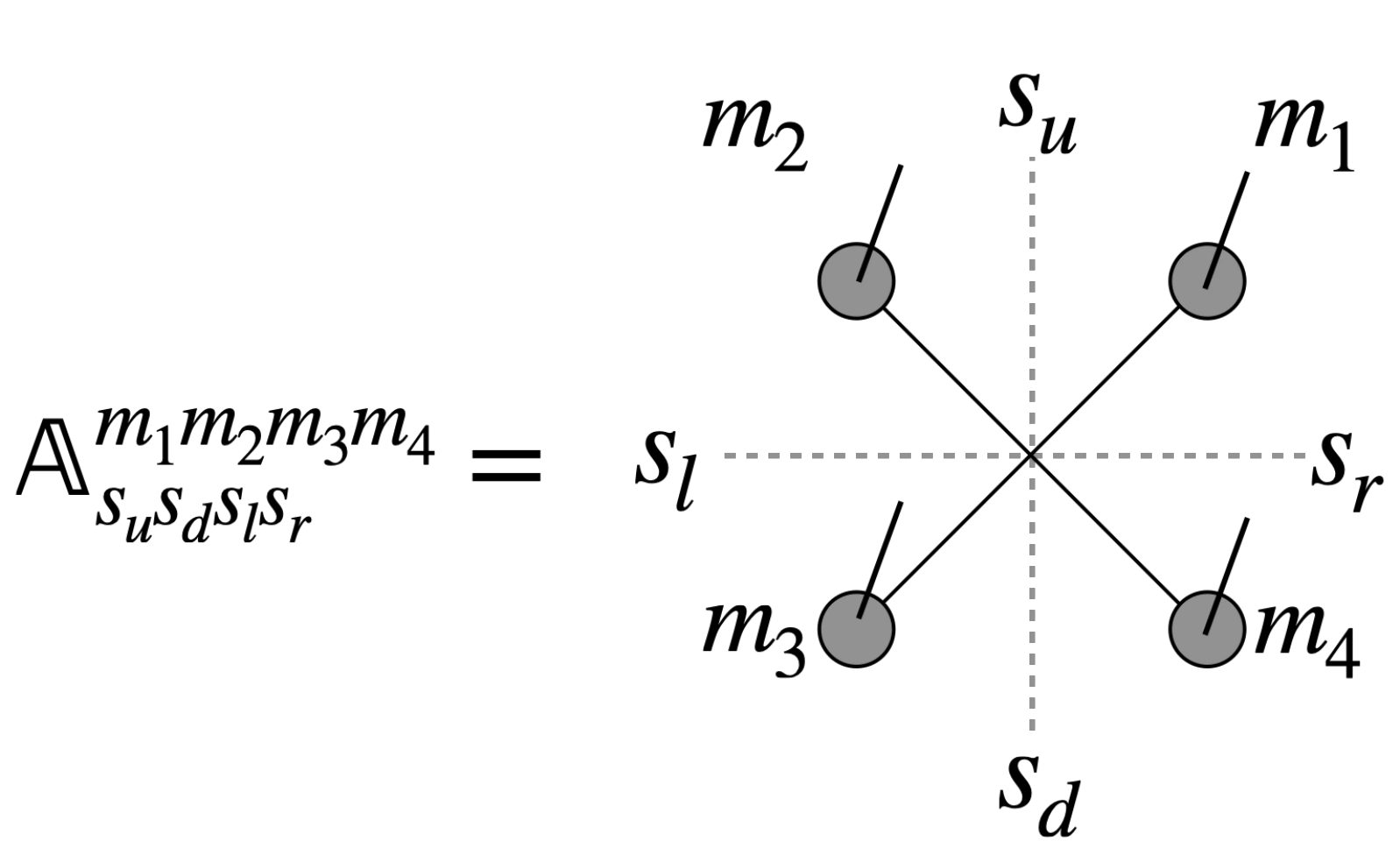}
    \label{eq:peps-z3tc-local}
\end{align}
The local tensor, \(\mathbb{A}_{s_u s_d s_l s_r}^{m_1 m_2 m_3 m_4}\), includes virtual indices \(s_u, s_d, s_l,\) and \(s_r\), and physical indices \(m_1, m_2, m_3,\) and \(m_4\). The virtual indices are internal degrees of freedom used for tensor contractions, while the physical indices represent local \(\mathbb{Z}_3\) spins. Each index takes values \(0, 1, 2\). The tensor element \(\mathbb{A}_{s_u s_d s_l s_r}^{m_1 m_2 m_3 m_4} = 1\) if the following conditions are satisfied:
\begin{align}
m_1 = & s_u - s_r  \mod 3, &
m_2 = & -s_u + s_l \mod 3, \nn
m_3 = & -s_d + s_l \mod 3, &
m_4 = &  s_d - s_r \mod 3.
\label{eq:peps-a}
\end{align}
Otherwise, \(\mathbb{A}_{s_u s_d s_l s_r}^{m_1 m_2 m_3 m_4} = 0\).

The tensor \(\mathbb{A}\) is designed to generate the closed-loop configurations of the \(\mathbb{Z}_3\) TC. The values of the virtual bonds correspond to the numbers on the plaquette crossed by the bond, indicating the number of times the \(b_p\) operator is applied to the plaquette. The physical bonds in Eq.\,\eqref{eq:peps-a} are assigned to represent the domain walls between neighboring plaquettes.

Operations on the physical indices of the tensor \(\mathbb{A}_{s_u s_d s_l s_r}^{m_1 m_2 m_3 m_4}\) can be expressed in terms of corresponding operations on the virtual indices:
\begin{align}
    \includegraphics[width=0.7\textwidth]{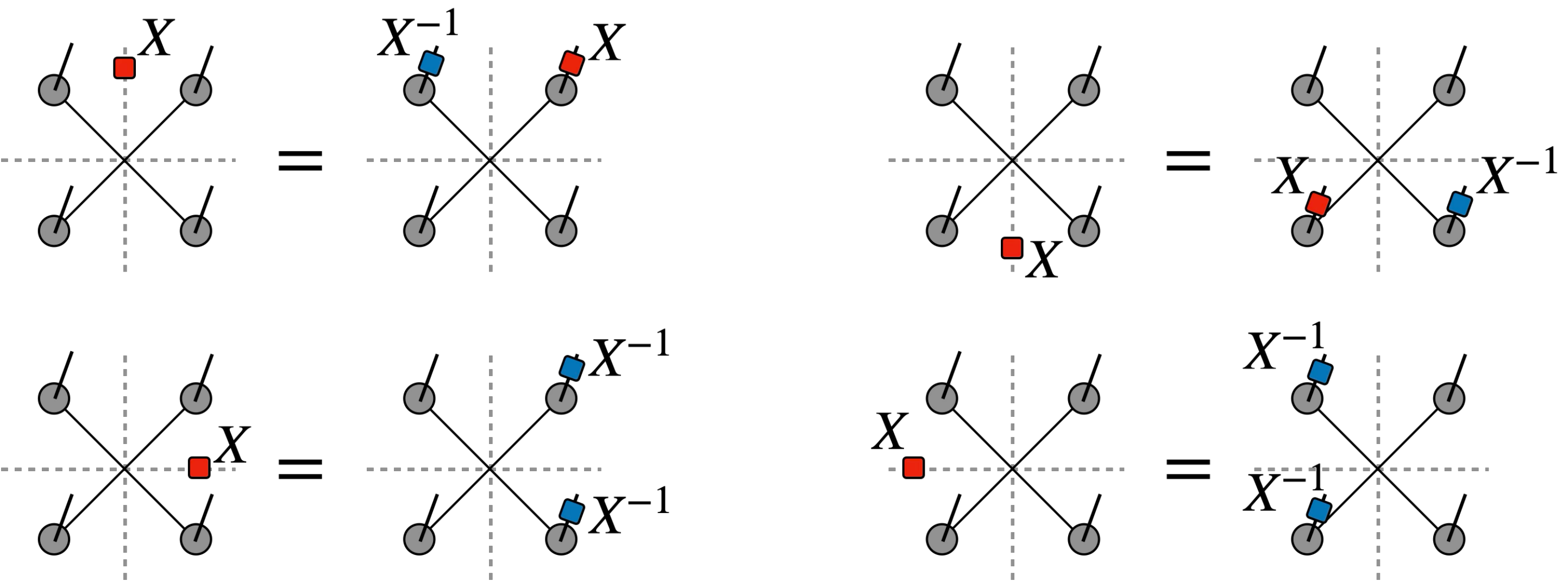}
    \label{eq:peps-z3tc-loc-x-sym}
\end{align}
and
\begin{align}
    \includegraphics[width=0.7\textwidth]{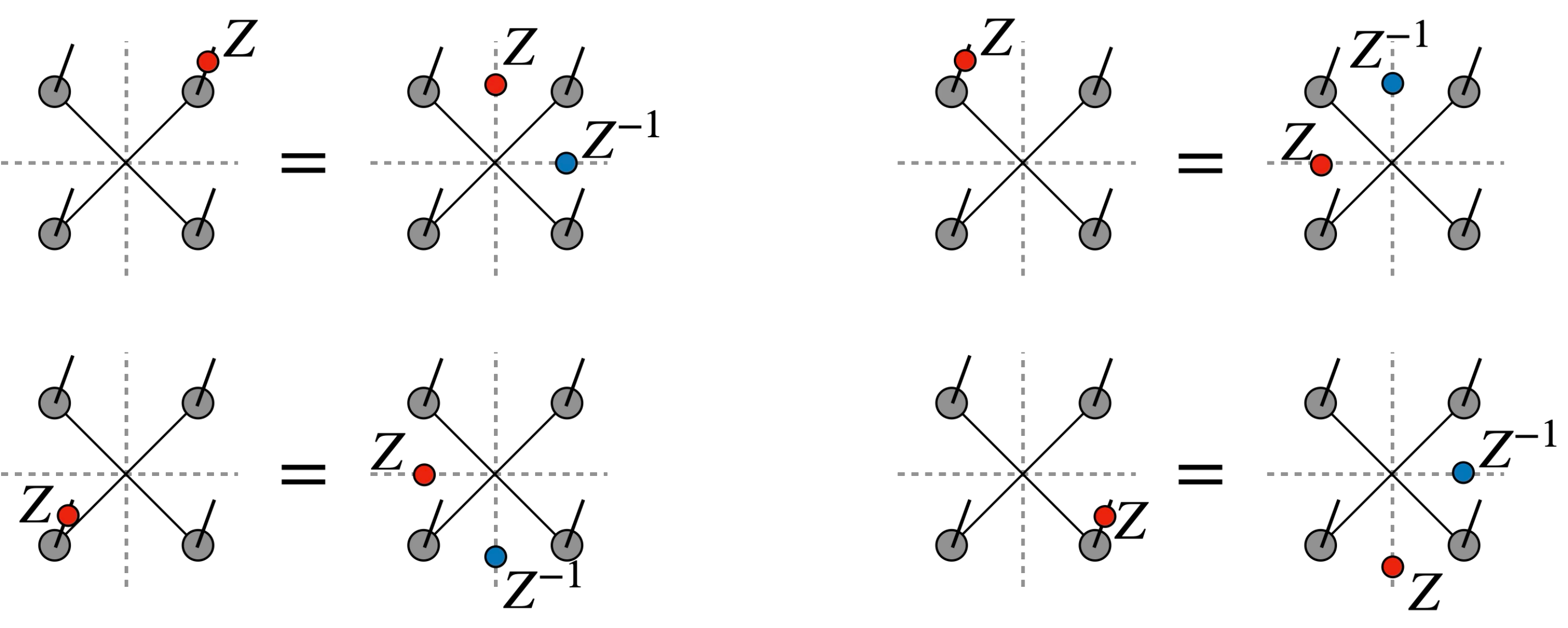}
    \label{eq:peps-z3tc-loc-z-sym}
\end{align}

In particular, Eq.\,\eqref{eq:peps-z3tc-loc-x-sym} defines the \(\mathbb{Z}_3\)-injective symmetry of the local tensor as:
\begin{align}
    \includegraphics[width=0.6\textwidth]{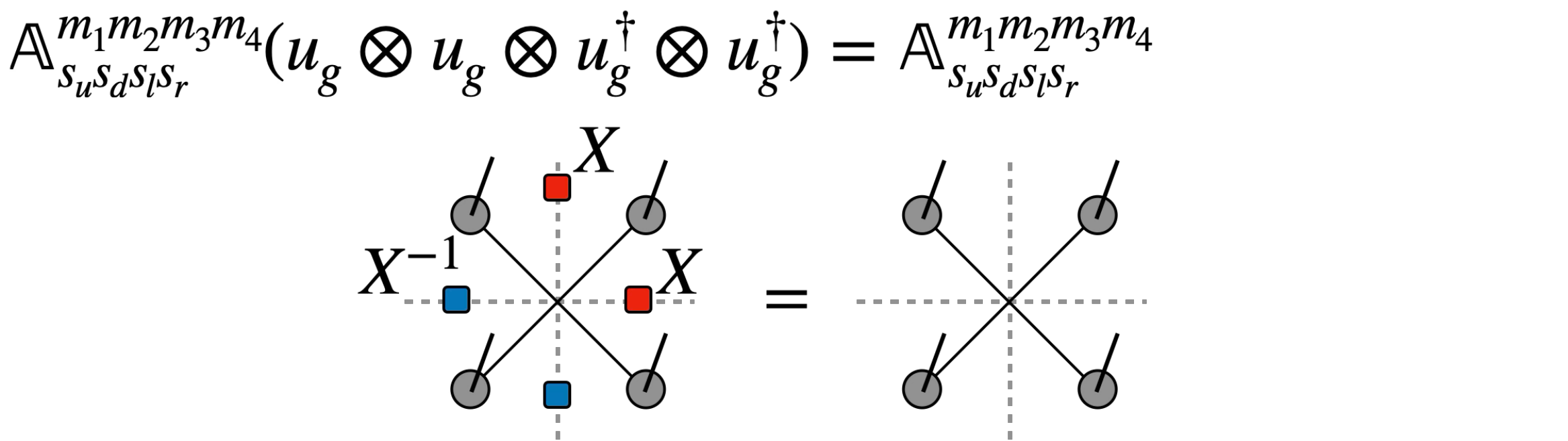}
    \label{eq:peps-z3tc-x-inj}
\end{align}
where \(u_g \in \{I, X, X^2\}\).

The PEPS representation for the deformed \(\mathbb{Z}_3\) TC is obtained by contracting the deformed local tensor, defined as:
\begin{align}
    {\mathbb{A}'}_{s_u s_d s_l s_r}^{m_1 m_2 m_3 m_4} = \mathbb{A}_{s_u s_d s_l s_r}^{m_1 m_2 m_3 m_4} D_{m_1}(\beta_x, \beta_z) D_{m_2}(\beta_x, \beta_z) D_{m_3}(\beta_x, \beta_z) D_{m_4}(\beta_x, \beta_z). \nonumber
\end{align}
Since the deformations are applied exclusively to the physical indices while leaving the virtual indices unchanged, the deformed tensor retains the same injective symmetry as described in Eq.\,\eqref{eq:peps-z3tc-x-inj}.

Using the PEPS representation, the overlap of the deformed wavefunction can be expressed as:
\begin{align}
    \includegraphics[width=0.5\textwidth]{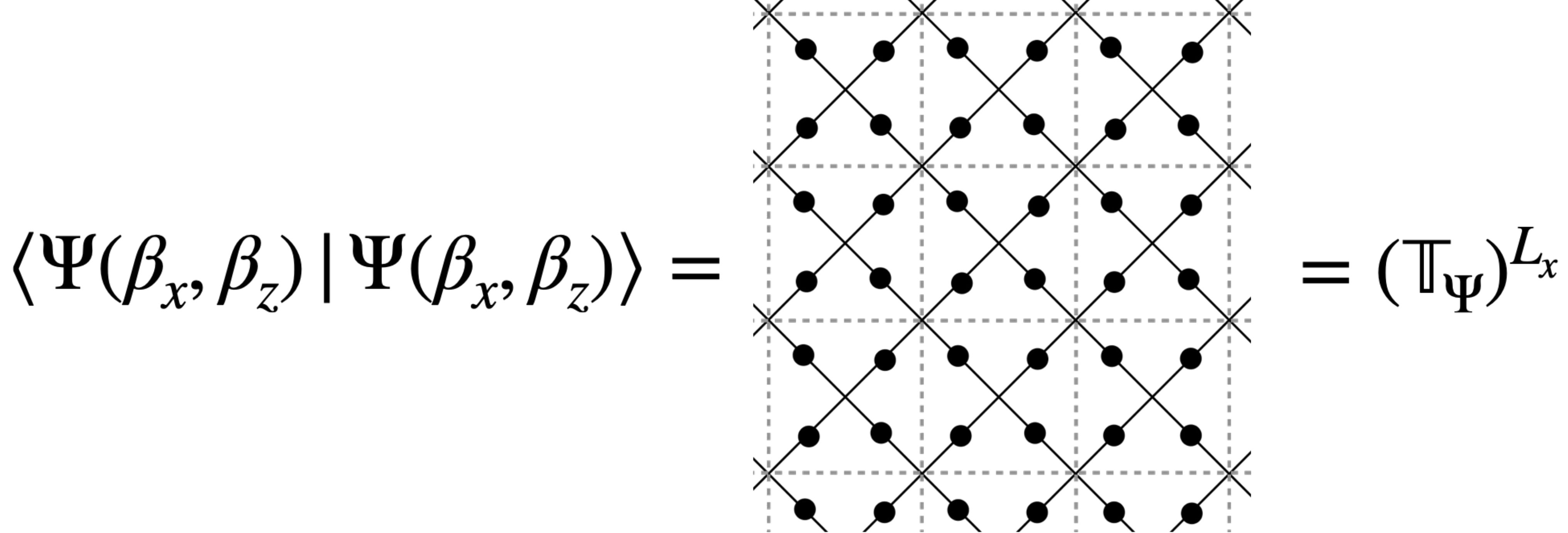}
    \label{eq:peps-z3tc-norm}
\end{align}
Here, \(\mathbb{T}_\Psi\) denotes the column-to-column transfer matrix, obtained by contracting the vertical virtual indices of the local tensors \(\mathbb{E}_\Psi\):
\begin{align}
    \includegraphics[width=0.8\textwidth]{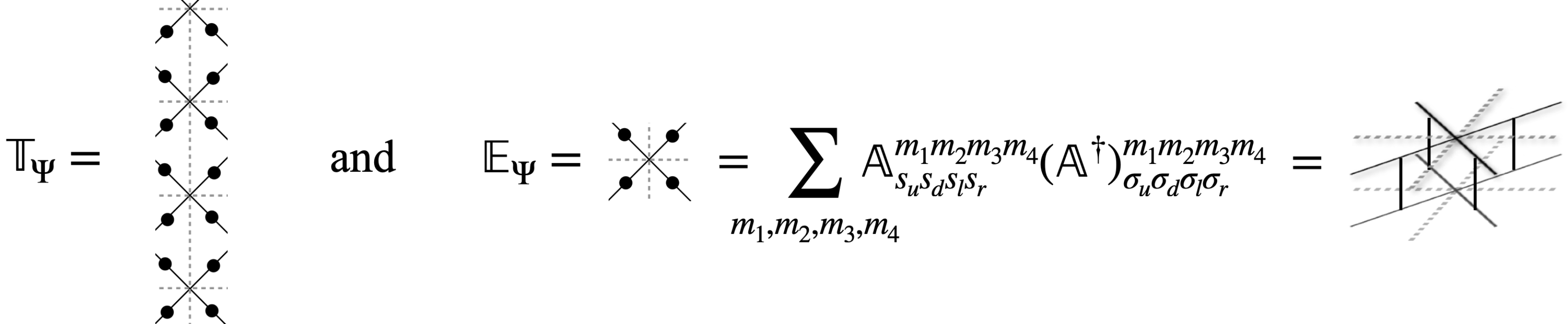}
    \label{eq:peps-z3tc-tm}
\end{align}

The injective symmetry of the local tensor \({\mathbb{A}'}_{s_u s_d s_l s_r}^{m_1 m_2 m_3 m_4}\) gives rise to two global \(\mathbb{Z}_3\) symmetries in the one-dimensional transfer matrix \(\mathbb{T}_\Psi\). By introducing a representation \(A_i \otimes B_i\), where \(A_i\) and \(B_i\) act on the \(i\)-th virtual bond of the transfer matrix in the ket and bra layers, respectively, the two global symmetries can be expressed as:
\begin{align}
{\cal U}_1 = \prod_i X_i \otimes X_i, 
\quad \text{and} \quad 
{\cal U}_2 = \prod_i I_i \otimes X_i.
\label{eq:tm-global-sym}
\end{align}
The corresponding order parameters, which indicate the spontaneous breaking of each symmetry, are given by:
\begin{align}
( Z_i \otimes I_i ),
\quad \text{and} \quad 
( Z_i \otimes Z_i ).
\end{align}

On the other hand, the classical partition function of the AT\(_3\) model, \(Z_{{\rm AT}_3}\), can also be expressed using a PEPS representation. This representation is obtained by contracting the local tensor:
\begin{align}
    \includegraphics[width=0.55\textwidth]{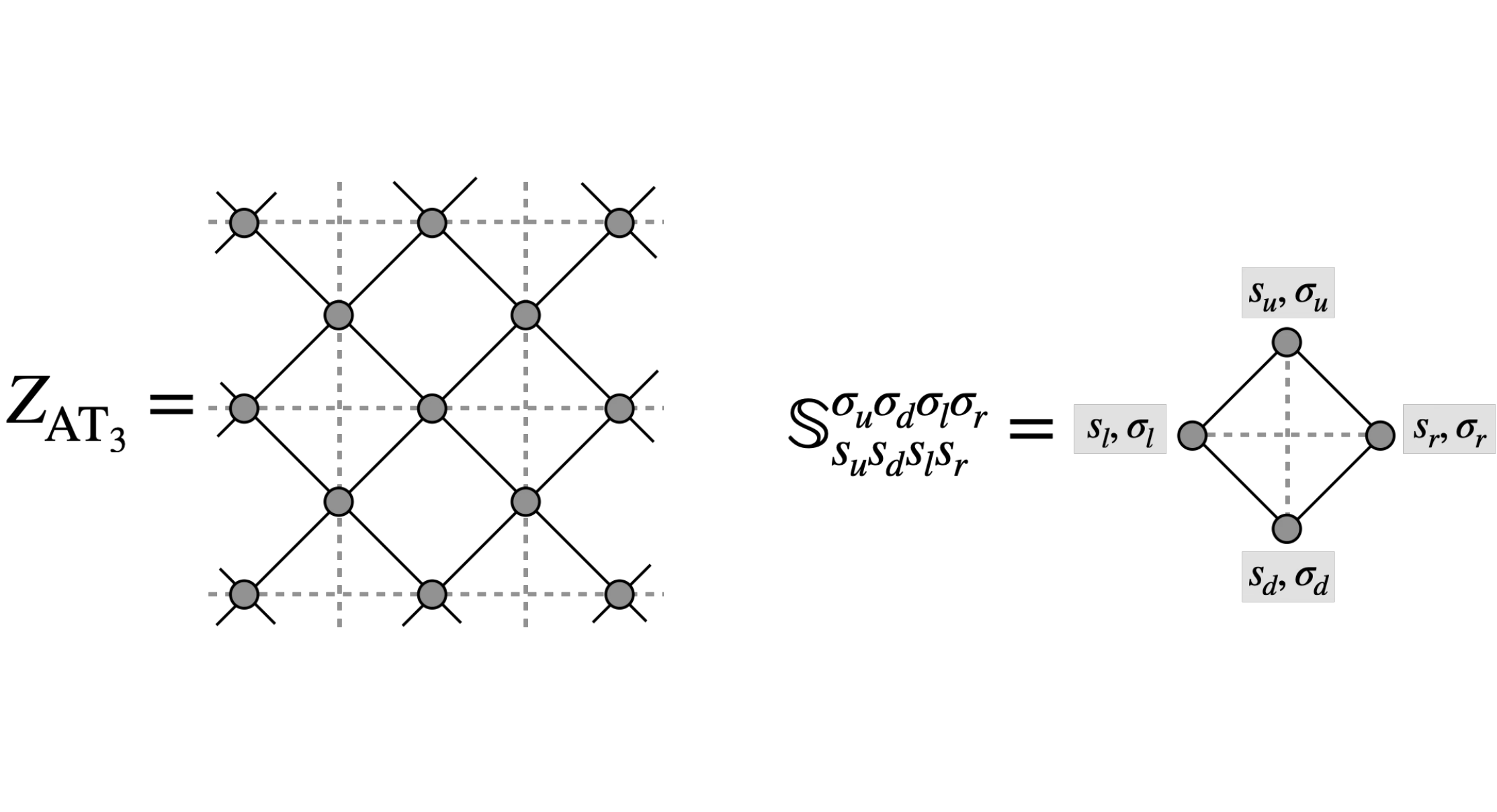}
    \label{eq:peps-at3}
\end{align}
where the local tensor \(\mathbb{S}_{s_u s_d s_l s_r}^{{\sigma_u \sigma_d \sigma_l \sigma_r}}\) is defined as:
\begin{align}
\log \mathbb{S}^{\sigma_u \sigma_d \sigma_l \sigma_r}_{s_u s_d s_l s_r} 
 =  &
  \beta_2 \left( \delta_{s_u, s_l} + \delta_{s_u, s_r} + \delta_{s_d, s_l} + \delta_{s_d, s_r} \right)
\nn
&
+ \beta_2 \left(  \delta_{\sigma_u, \sigma_l} + \delta_{\sigma_u, \sigma_r} + \delta_{\sigma_d, \sigma_l} + \delta_{\sigma_d, \sigma_r}  \right)
\nn
& 
+ \beta_{4,1} \big( 
\delta_{s_u - \sigma_l, s_l - \sigma_u } + 
\delta_{s_u - \sigma_r, s_r - \sigma_u } + 
\delta_{s_d - \sigma_d,  s_l- \sigma_l} + 
\delta_{s_d - \sigma_d , s_r- \sigma_r} 
\big)
\nn
& 
+ \beta_{4,2} \big( 
\delta_{s_u + s_l, \sigma_u + \sigma_l} + 
\delta_{s_u + s_r, \sigma_u + \sigma_r} + 
\delta_{s_d + s_l, \sigma_d + \sigma_l} + 
\delta_{s_d + s_r, \sigma_d + \sigma_r} 
\big),
\end{align}
This tensor is designed to represent the domain wall excitations between sites in Eq.\,\eqref{eq:at-z3tc-map}.

Since both local tensors \(\mathbb{E}_{\Psi}\) and \(\mathbb{S}\) are constructed to represent the domain walls between the numbers assigned to neighboring virtual bonds, they are inherently related. A closer examination reveals that the two tensors \(\mathbb{E}_{\Psi}\) and \(\mathbb{S}\) are equivalent under the mapping provided in Eq.\,\eqref{eq:at-z3tc-map}. Consequently, the column-to-column transfer matrix \(\mathbb{T}_{{\rm AT}_3}\), formed by contracting the vertical virtual bonds of \(\mathbb{S}\), is also symmetric under the operations \({\cal U}_1\) and \({\cal U}_2\) given in Eq.\,\eqref{eq:tm-global-sym}.

Since the PEPS representation of the classical partition function \(Z_{{\rm AT}_3}\) in Eq.\,\eqref{eq:peps-at3} can be expressed as a repeated product of the transfer matrix \(\mathbb{T}_{{\rm AT}_3}\), the two global symmetries \({\cal U}_1\) and \({\cal U}_2\) of the transfer matrix naturally correspond to the two global symmetries of the AT\(_3\) model, \(U_1\) and \(U_2\), as defined in Eq.\,\eqref{eq:at3-global-sym}.

\subsection{Phase diagram}
\label{sec:tn-phase-diagram}

\begin{figure}[ht]
\centering
\setlength{\abovecaptionskip}{5pt}
\includegraphics[width=0.5\textwidth]{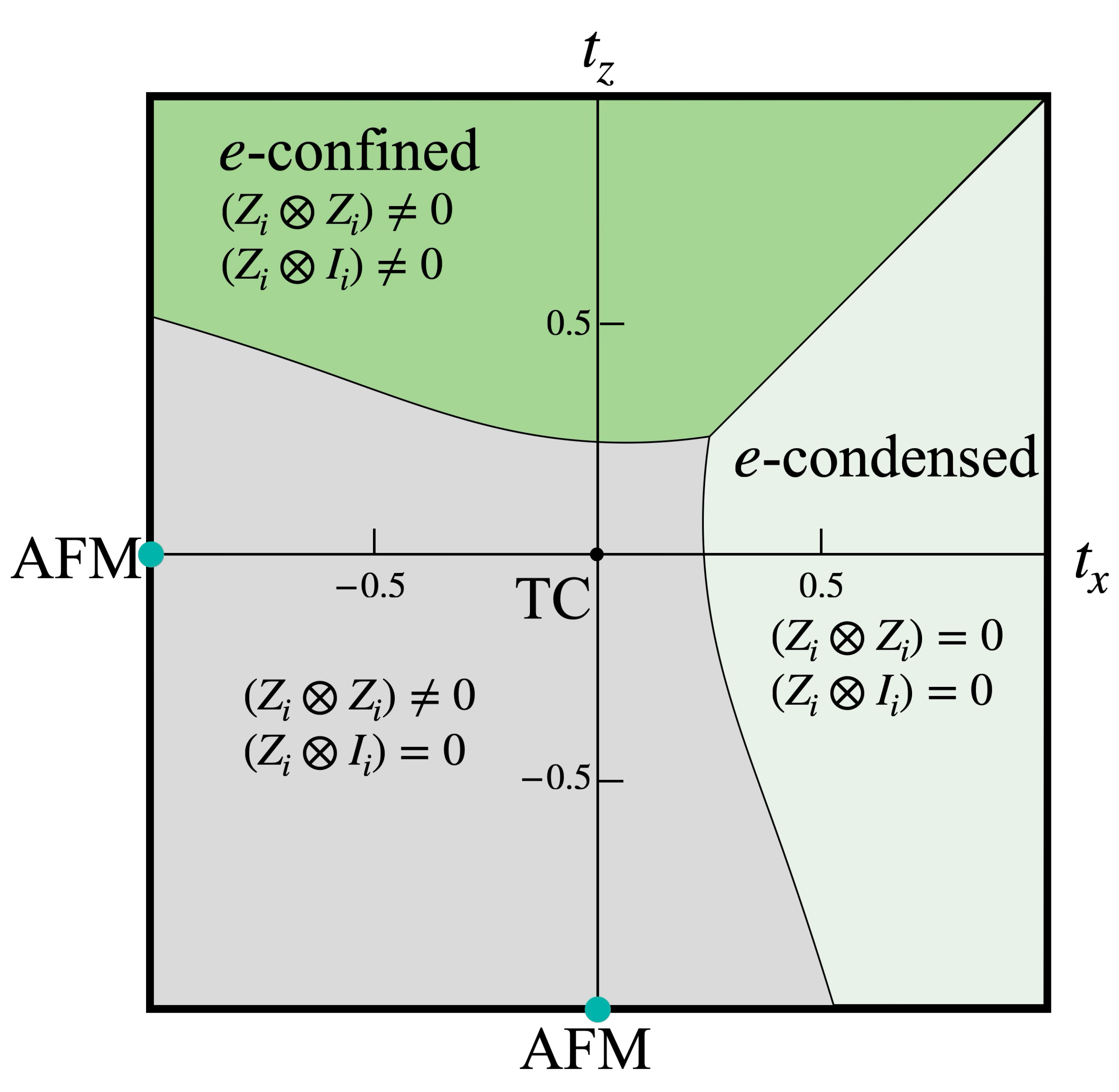}
\caption{Phase diagram of the deformed $\mathbb{Z}_3$ toric code as a function of the variational parameters $t_x = \tanh(\beta_x/2)$ and $t_z = \tanh(\beta_z/2)$, obtained with the \blue{variational uniform matrix-product-state (VUMPS) method}. Three phases are identified: the TC phase (gray, $(Z_i \otimes Z_i) \neq 0$, $(Z_i \otimes I_i) = 0$), the $e$-confined phase (dark green, both order parameters nonzero), and the $e$-condensed phase (light green, both order parameters zero). The phase diagram is symmetric about the $t_z = t_x$ line due to the $\mathcal{D}_{\rm e\text{-}m}$ duality. Two isolated \blue{antiferromagnetic (AFM)} critical points (cyan dots) with $c=1$ are located at $(t_x, t_z) = (-1, 0)$ and $(0, -1)$.
} \label{fig:z3tc-phase-diagram}
\end{figure}

Recall that the norm of the deformed \(\mathbb{Z}_3\) TC (\(\langle \tilde{\Psi}(\beta_x, \beta_z) | \tilde{\Psi}(\beta_x, \beta_z) \rangle\)) or the classical partition function of the AT\(_3\) model (\(Z_{{\rm AT}_3}\)) can be expressed as a repeated product of the column-to-column transfer matrix \(\mathbb{T}_\Psi\) or \(\mathbb{T}_{{\rm AT}_3}\), respectively. This involves applying the transfer matrix \(\mathbb{T}\) \(L_x\) times to the far-right boundary state \(|\varphi_r^0)\) and taking the overlap with the far-left boundary state \((\varphi^0_l|\). Specifically, these quantities can be written as:
\begin{align}
\langle \tilde{\Psi}(\beta_x, \beta_z) | \tilde{\Psi}(\beta_x, \beta_z) \rangle &= (\varphi^0_l | \left[\mathbb{T}_\Psi\right]^{L_x} |\varphi^0_r), \\
Z_{{\rm AT}_3} &= (\varphi^0_l | \left[\mathbb{T}_{{\rm AT}_3}\right]^{L_x} |\varphi^0_r).
\end{align}

Now, consider a right boundary state evolved by applying the transfer matrix \(x\) times: \( |\varphi^x_r) = \left[\mathbb{T}\right]^{x} |\varphi_r^0)\). In the thermodynamic limit, this state satisfies the relation \( |\varphi_r^{x+1}) = \mathbb{T} |\varphi_r^x) \propto |\varphi_r^x)\). This implies that the right boundary state \(|\varphi_r^x)\) in the thermodynamic limit corresponds to the eigenstate of the transfer matrix \(\mathbb{T}\) associated with the largest eigenvalue. Eigenstates with smaller eigenvalues vanish under repeated applications of \(\mathbb{T}\) due to normalization effects. 

Using the VUMPS method\,\cite{fishman18prb}, we numerically compute the fixed-point eigenstate of the transfer matrix \(\mathbb{T}\). In Ref.\,\cite{haegeman15}, it is demonstrated that the spontaneous symmetry breaking of the two global symmetries of the transfer matrix, \({\cal{U}}_1\) and \({\cal{U}}_2\), is directly connected to the confinement and condensation of \(e\) anyons. The phase diagram of the deformed toric code, derived from the analysis of the order parameters \((Z_i \otimes I_i)\) and \((Z_i \otimes Z_i)\), is shown in Fig.\,\ref{fig:z3tc-phase-diagram}, where the \(x\)- and \(y\)-axes represent the deformation parameters \(t_x\) and \(t_z\), respectively. 

The TC point lies in a phase characterized by \((Z_i \otimes Z_i) \neq 0\) and \((Z_i \otimes I_i) = 0\). Here, \((Z_i \otimes Z_i) \neq 0\) indicates that the \(e\) anyons are not condensed, while \((Z_i \otimes I_i) = 0\) indicates that the \(e\) anyons are deconfined. As \(t_x\) increases, the system transitions to a fully symmetric phase, where \((Z_i \otimes Z_i) = 0\), indicating that the \(e\) anyons are condensed. Conversely, as \(t_z\) increases, the system transitions to a fully symmetry-broken phase, where \((Z_i \otimes I_i) \neq 0\), signifying that the \(e\) anyons are confined.

In the context of the classical AT\(_3\) model, the spontaneous symmetry breaking of the fixed point \(|\varphi_r^x)\) corresponds directly to the symmetry breaking of the global symmetries of the AT\(_3\) system. The TC phase, the \(e\)-confined phase, and the \(e\)-condensed phase correspond to the partially ordered phase, the fully ordered FM phase, and the paramagnetic phase of the classical AT\(_3\) model, respectively. These phases are also present in the \(\mathbb{Z}_2\) version of the original Ashkin--Teller model. 

\begin{revision}
The isolated points $(t_x,t_z)=(-1,0)$ and $(0,-1)$ inherit the $c=1$ AFM
endpoint of the $Q=3$ Potts model and are a concrete qualitative distinction
from the $\mathbb Z_2$ wavefunction phase diagram reviewed in the Appendix.
The AT$_3$ construction has three couplings, while the wavefunction studied
here realizes a two-parameter slice; the unexplored coupling direction is left
as a separate classical-model problem rather than used to infer additional
phases in the present diagram.
\end{revision}

\begin{figure}[ht]
\centering
\setlength{\abovecaptionskip}{5pt}
\includegraphics[width=0.93\textwidth]{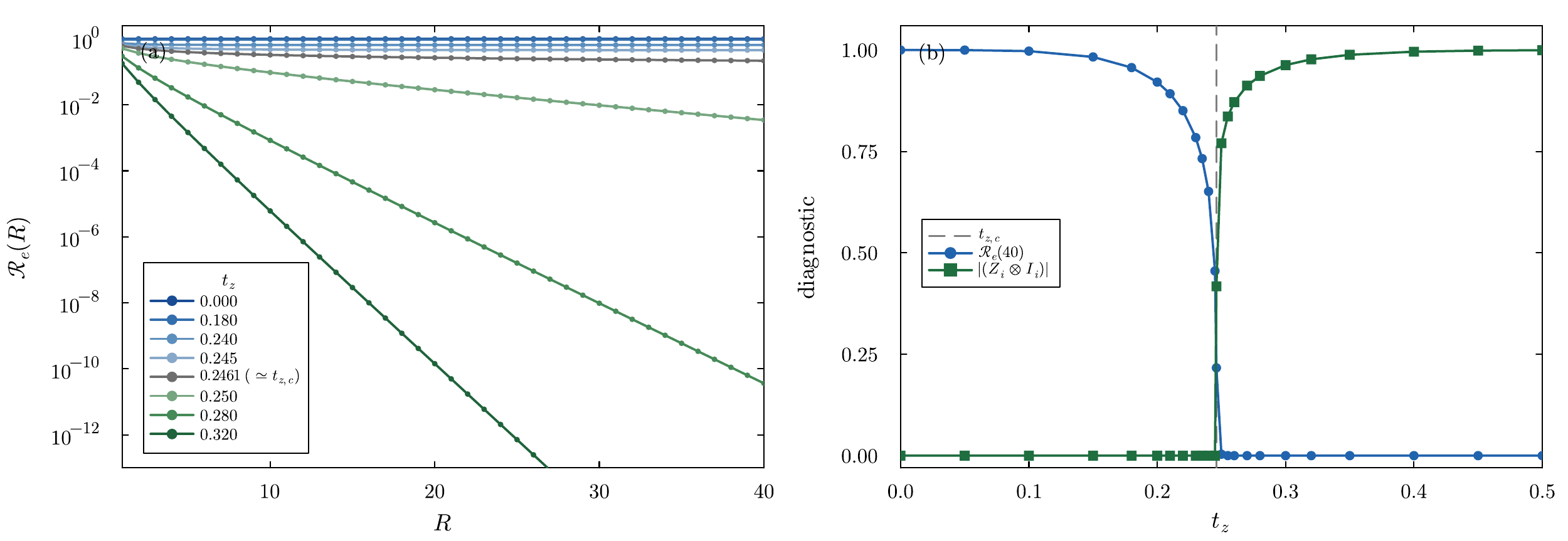}
\caption{{Fredenhagen--Marcu-type finite-distance line-tension diagnostic on the
$t_x=0$ cut. (a) The normalized $e$-anyon pair-state norm $\mathcal R_e(R)$
versus separation for representative $t_z$. It saturates in the TC phase and
decays exponentially in the $e$-confined phase. (b) $\mathcal R_e(40)$ and the
transfer-matrix order parameter $|(Z_i\otimes I_i)|$. The dashed line marks
the analytic Potts value
$t_{z,c}=\tanh[\log(1+\sqrt3)/4]\simeq0.2461$. The calculation uses bond
dimension $\chi=30$ and $R_{\max}=40$; spot checks at $\chi=40$ near the
transition agree at the shown resolution.}}
\label{fig:fm-type-line-tension}
\end{figure}

\begin{revision}
As an independent finite-distance check, we evaluated $\mathcal R_e(R)$ along
the $t_x=0$ cut using the same norm transfer matrix. In the TC phase it
approaches a nonzero, separation-independent value, whereas in the
$e$-confined phase it decays exponentially with $R$
[Fig.~\ref{fig:fm-type-line-tension}(a)]. At the largest computed separation,
$R=40$, its collapse agrees within the resolution of the sampled cut and finite
bond dimension with both the onset of $|(Z_i\otimes I_i)|$ and the analytic
Potts boundary $t_{z,c}\simeq0.2461$
[Fig.~\ref{fig:fm-type-line-tension}(b)]. This supplies a finite-distance
consistency check of the confinement boundary; it is not used to refit or
redefine the phase diagram.
\end{revision}

\FloatBarrier

\subsection{\texorpdfstring{\blue{Topological entanglement entropy}}
{Topological entanglement entropy}}
\label{sec:tee}

\begin{revision}
The gapped TC region realizes the Abelian quantum double $D(\mathbb Z_3)$.
It has nine anyon types, each with quantum dimension $d_a=1$, and therefore
total quantum dimension
\begin{align}
\mathcal D=\sqrt{\sum_a d_a^2}=3.
\end{align}
The universal constant in the bipartite entanglement entropy, known as the
topological entanglement entropy (TEE), is consequently
\begin{align}
\gamma=\log\mathcal D=\log3
\label{eq:tee-z3}
\end{align}
~\cite{kitaev06tee,levin06tee}. Equivalently, for this Abelian quantum double
only, the torus ground-state degeneracy (GSD) is $9$ and
$\gamma=\tfrac12\log({\rm GSD})$.

This value is consistent with the tensor-network information already used
above. The deformation acts on physical indices while preserving the
$\mathbb Z_3$-injective virtual symmetry, and the transfer-matrix fixed point
retains the topological symmetry realization throughout the gapped TC region
~\cite{schuch10,cirac11,schuch13,haegeman15}. The virtual symmetry and
transfer-matrix fixed point obtained in our calculation therefore provide the
same phase-level topological distinction as the TEE: $\gamma=\log3$ throughout
the gapped TC region, whereas the gapped condensed and confined regions are
topologically trivial and have $\gamma=0$. At a critical boundary the usual
gapped-phase constant-term interpretation need not apply, so we assign no
plateau value to the critical point itself.
\end{revision}

\subsection{Criticalities}
\label{sec:criticalities}

\begin{figure}[ht]
\centering
\setlength{\abovecaptionskip}{5pt}
\includegraphics[width=0.8\textwidth]{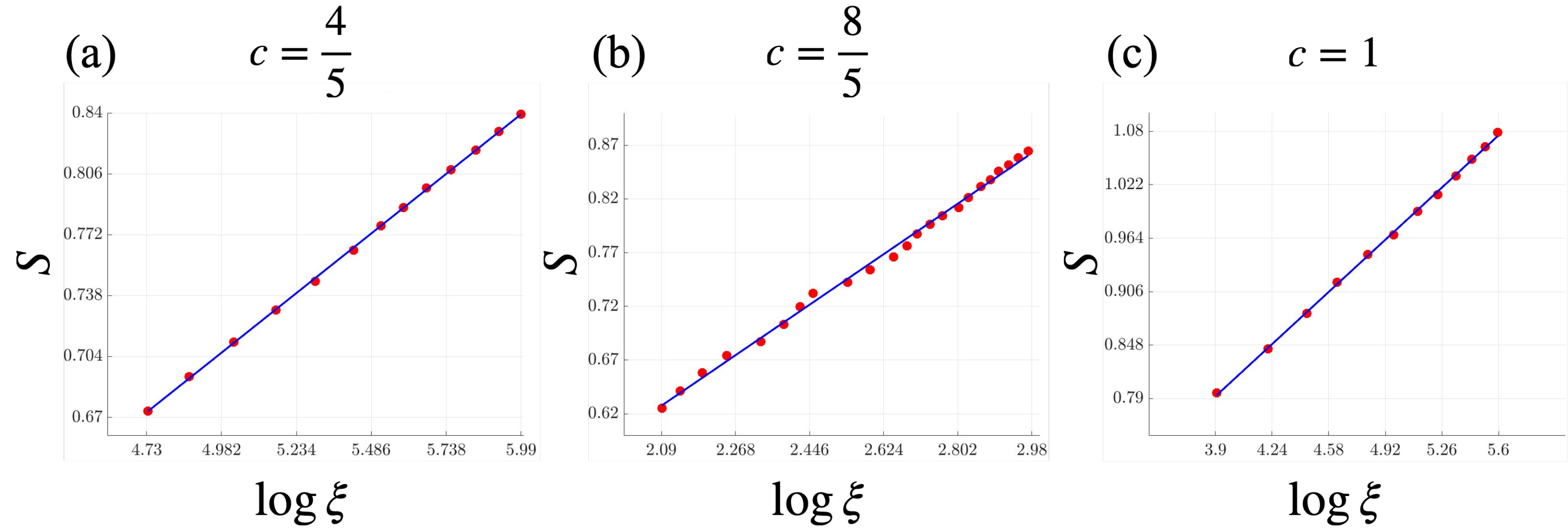}
\caption{Central charge determination at the critical points from the Calabrese--Cardy scaling of entanglement entropy $S$ versus the logarithm of the correlation length $\log \xi$ of the \blue{variational uniform matrix-product-state (VUMPS)} fixed-point state. (a) The phase boundary between the \blue{toric-code (TC)} and $e$-condensed (or $e$-confined) phases yields $c = 4/5$, corresponding to the $\mathbb{Z}_3$ parafermion \blue{conformal field theory (CFT)}. (b) The merged critical line separating the $e$-confined and $e$-condensed phases gives $c = 8/5$. (c) The isolated \blue{antiferromagnetic (AFM)} critical points exhibit $c = 1$, consistent with the $\mathbb{Z}_4$ parafermion CFT.
} \label{fig:central-charge}
\end{figure}

% \subsection{Criticalities}

We analyzed the central charges at the critical points by fitting the entanglement entropy and correlation length of the fixed point \(|\varphi_r^x)\) to the Calabrese--Cardy formula, as illustrated in Fig.\,\ref{fig:central-charge}. The phase boundaries between the TC phase and the \(e\)-condensed or \(e\)-confined phases contain critical points described by the \(Q=3\) Potts model, which corresponds to a \(\mathbb{Z}_3\) parafermion CFT. The central charge \(c = \frac{4}{5}\) at these critical points is confirmed, as shown in Fig.\,\ref{fig:central-charge}(a). 

The two phase boundaries merge into a single critical line dividing the \(e\)-confined and \(e\)-condensed phases. Along this line, the central charge is \(c = \frac{8}{5}\), as depicted in Fig.\,\ref{fig:central-charge}(b). Lastly, two isolated antiferromagnetic (AFM) critical points are identified within the \(e\)-confined and \(e\)-condensed phases, respectively. These points exhibit a central charge of \(c = 1\), consistent with the \(\mathbb{Z}_4\) parafermion CFT, as depicted in Fig.\,\ref{fig:central-charge}(c).

\subsection{\texorpdfstring{\blue{Qualitative distinctions from the
$\mathbb Z_2$ case}}{Qualitative distinctions from the Z2 case}}
\label{sec:z3-distinctions}

\begin{revision}
\paragraph{Absence of sign-change folding.}
For qubits, the anticommutation of $X$ and $Z$ produces unitary relations that
fold the deformed-$\mathbb Z_2$ wavefunction diagram under sign changes of its
parameters. For qutrits, $ZX=\omega XZ$ does not yield the same folding. Its
visible consequence in the present wavefunction family is that the
$t_z=-1$ and $t_x=-1$ AFM endpoints remain as distinct, accessible boundary
points rather than being identified with their positive-parameter images.

\paragraph{Independent four-spin structures.}
The AT$_3$ norm contains two three-state Potts variables and two independent
four-spin constraints, associated with their sum and difference modulo three.
For $\mathbb Z_2$ those constraints coincide because sum and difference are
identical modulo two, leaving the single four-spin structure of the standard
Ashkin--Teller model. This algebraic construction also differs from the
$N$-color Ashkin--Teller generalization, which couples multiple Ising colors
symmetrically~\cite{kohmoto81,grest81}. The current wavefunction explores a
two-parameter slice of the resulting three-coupling AT$_3$ space.

\paragraph{Critical and constrained structures.}
The $c=4/5$ and $c=8/5$ critical lines, the isolated $c=1$ AFM endpoints, and
the square-ice limit with an emergent $U(1)$ one-form symmetry provide the
specific content of the $\mathbb Z_3$ phase diagram. At the square-ice point,
the same constraint structure produces Hilbert-space fragmentation and the
exact count $2^{L+2}-4$ of unflippable scar configurations.

\paragraph{Platform context.}
A $\mathbb Z_3$ toric-code ground state and its defects have recently been
prepared on 24 encoded qutrits in a trapped-ion processor
~\cite{iqbal25qutrit}. This result establishes direct hardware relevance for
qutrit topological-code wavefunctions. The deformed-state phase diagram studied
here is theoretical, however, and we do not claim an experimental realization
of the deformed family.
\end{revision}

\section{Conclusion}
\label{sec:conclusion}

\begin{revision}
We have studied phase transitions in a measurement-prepared family of
deformed $\mathbb Z_3$ toric-code wavefunctions. Starting from the cluster
state on the Lieb lattice, the construction produces the filtered states
$|\tilde\Psi(\beta_x)\rangle$ and their electric--magnetic dual
$|\tilde\Psi(\beta_z)\rangle$. We derived the dressed local stabilizer
constraints and a Hermitian frustration-free parent Hamiltonian for the
filtered state. This parent construction establishes the exact wavefunction;
the phase diagram itself is determined by the two-dimensional norm problem,
not by interpreting the parent as a field-driven thermodynamic Hamiltonian.

For a single $\beta_z$ deformation, the norm is weighted by
$e^{-2\beta_zl_N}$ and maps to the $Q=3$ Potts model with
$\beta=2\beta_z$. The TC-to-$e$-confined transition therefore occurs at
\begin{align}
e^{2\beta_{z,c}}=1+\sqrt3,
\qquad
e^{\beta_{z,c}}=\sqrt{1+\sqrt3},
\end{align}
with $c=4/5$; the dual $\beta_x$ cut gives the corresponding condensation
transition. The finite-distance diagnostic $\mathcal R_e(R)$ independently
saturates in the TC phase and decays exponentially in the confined phase,
consistent with this boundary. At $\beta_z\to-\infty$ the wavefunction reaches
the square-ice ensemble with its emergent $U(1)$ one-form charge,
fragmentation, and exact unflippable scar configurations.

For the two-parameter family, the norm realizes a two-parameter slice of the
three-coupling AT$_3$ construction, whose two independent $\mathbb Z_3$
symmetries organize the TC, $e$-confined, and $e$-condensed phases. VUMPS
calculations yield critical structures with $c=4/5$, $c=8/5$, and isolated
$c=1$ AFM endpoints. In the gapped TC region, the $D(\mathbb Z_3)$ anyon data,
virtual $\mathbb Z_3$-injective symmetry, and transfer-matrix fixed-point
structure give $\mathcal D=3$ and $\gamma=\log3$; the gapped condensed and
confined phases are topologically trivial and have $\gamma=0$.

The substantive distinctions from the qubit problem are the absence of the
sign-change folding, the independence of the two four-spin structures modulo
three, the isolated AFM endpoints, and the square-ice one-form
constraint structure. Natural follow-up problems include the remaining AT$_3$
coupling direction, the constrained dynamics at the square-ice endpoint, and
extensions to other quantum-double wavefunctions. The present results are
theoretical; the recent qutrit toric-code experiment motivates their platform
relevance but is not an experimental realization of the deformed family.
\end{revision}

\section*{Acknowledgments}

\paragraph{Funding information}
This work was supported by the Basic Science Research Program through the National Research Foundation of Korea funded by the Ministry of Science and ICT [Grant No.\ RS-2023-00220471, RS-2025-16064392].

\begin{appendix}
\numberwithin{equation}{section}

\section{Review: Deformed $\mathbb{Z}_2$ Toric Code}
\label{sec:z2-review}

\begin{figure}[ht]
\centering
\setlength{\abovecaptionskip}{5pt}
\includegraphics[width=0.35\textwidth]{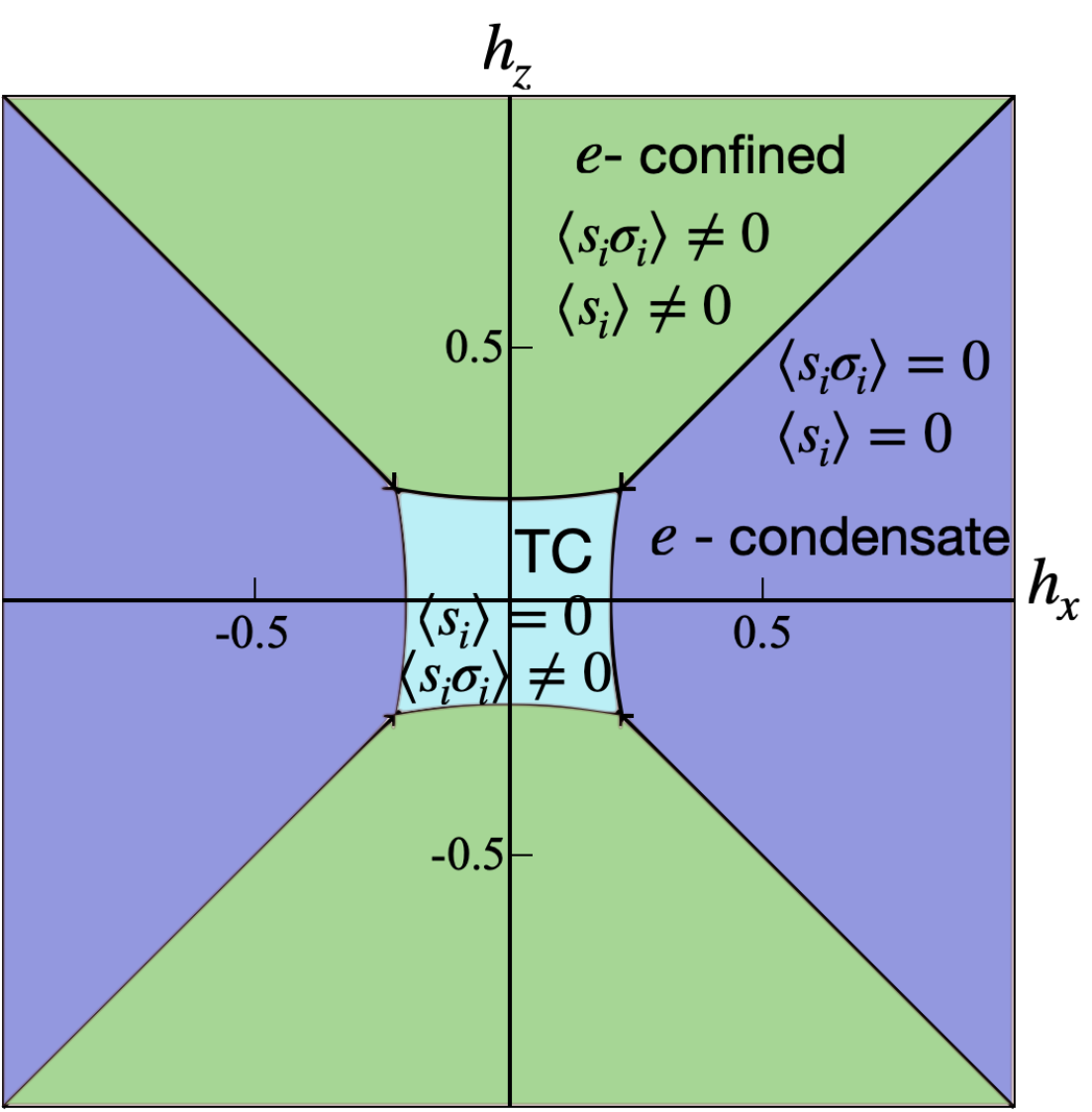}
\caption{
The phase diagram of the deformed $\mathbb{Z}_2$ toric code as a function of the variational parameters $h_x$ and $h_z$. The order parameters of the corresponding classical AT phases, $\langle s_i \rangle$ and $\langle s_i \sigma_i \rangle$, are indicated.
} \label{fig:z2-phase-diagram}
\end{figure}

In this appendix, we review the approach presented in Ref.\,\cite{zhu-19prl}, where the phase diagram of the deformed $\mathbb{Z}_2$ toric code wavefunction is derived by mapping its norm to the partition functions of classical models. In certain limits of the deformation parameters, the norm of the deformed wavefunction maps directly to the partition function of the classical Ising model. The condensation and confinement of electric ($e$) anyons are closely tied to the symmetry-breaking transitions of the corresponding classical Ising model. In the more general parameter regime, the norm of the wavefunction maps to the partition function of the Ashkin--Teller (AT) model~\cite{kohmoto81, saleur88, caselle07, aoun23}. The phases of the Ising model generalize into three distinct phases of the AT model, which are characterized by two order parameters. These three phases correspond to the three phases of the deformed $\mathbb{Z}_2$ toric code wavefunction, as illustrated in Fig.\,\ref{fig:z2-phase-diagram}.

\subsection{Loop-gas picture}
\label{sec:z2-loop-gas}

\begin{figure}[ht]
\centering
\setlength{\abovecaptionskip}{5pt}
\includegraphics[width=0.8\textwidth]{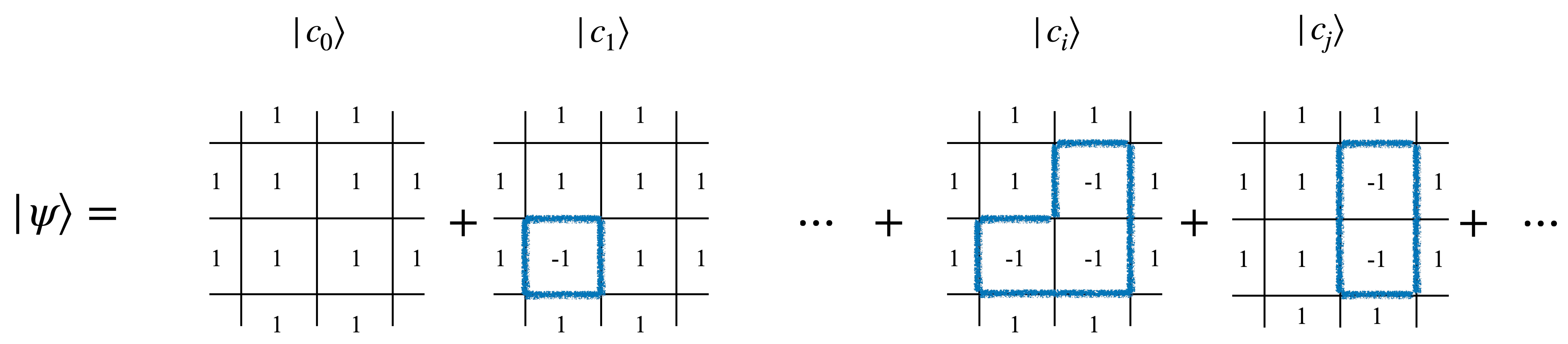}
\caption{{Bare loop-gas representation of the undeformed
$\mathbb Z_2$ toric-code ground state
$|{\rm TC}_2\rangle=\sum_c|c\rangle$. The blue lines denote closed loops
separating plaquettes with different stabilizer labels; occupied and empty
links carry $|1\rangle$ and $|0\rangle$, respectively.}}
\label{fig:z2-loop-gas}
\end{figure}

We analyze the phases of the deformed toric code by examining the wavefunction norm through its mapping to a classical model. Each term in the expansion of Eq.\,(\ref{eq:toric-code-gs}) corresponds to a closed-loop configuration, as illustrated in \blue{Fig.\,\ref{fig:z2-loop-gas}}. The number at the center of a plaquette indicates whether the stabilizer $b_p$ is applied ($-1$) or not ($1$). The loops act as domain walls separating regions labeled by $1$ and $-1$, and these loops are always closed. On the links, an occupied link represents the local state $|1\rangle$, while an empty link corresponds to $|0\rangle$.

The bare ground state wavefunction, without deformation, is an equal superposition of all possible closed-loop configurations:
\begin{align}
|{\rm TC}_2\rangle = \sum_c |c\rangle,
\label{eq:z2-loop-gas-app}
\end{align}
where the summation runs over all possible closed-loop configurations $c$, which form an orthonormal basis.

In the loop-gas representation, the loops are closely tied to $e$ anyons. Each anyon corresponds to the endpoint of a loop, where $a_v = -1$. In the ground state, only closed-loop configurations are present, ensuring $a_v = 1$ at every vertex. Anyons are created in pairs by ``cutting'' a loop, introducing open loops with $e$ anyons at the endpoints.

\subsection{Wavefunction deformation}

The deformed $\mathbb{Z}_2$ toric code wavefunction is expressed as:
\begin{align}
|\psi(h_x, h_z)\rangle
= P_2(h_x, h_z) |{\rm TC}_2\rangle
= \sum_c |\tilde{c}\rangle,
\label{eq:z2tc-deformed-wfn}
\end{align}
where the deformation operator $P_2(h_x, h_z) \equiv \prod_l p_{2,l}(h_x, h_z)$ with
\begin{align}
p_{2,l}(h_x, h_z) = \mathbb{I}_{2,l} + h_x X_l + h_z Z_l.
\end{align}
Here, $X$ and $Z$ are $\mathbb{Z}_2$ Pauli operators. The deformation operator rotates and adjusts the norm of the local states:
\begin{align}
|0\rangle \rightarrow
|\tilde{0}\rangle =
\begin{pmatrix}
1 + h_z \\ h_x
\end{pmatrix}, \quad
|1\rangle \rightarrow
|\tilde{1}\rangle =
\begin{pmatrix}
h_x \\
1 - h_z
\end{pmatrix}.
\end{align}
Setting $\langle \tilde{0} | \tilde{0} \rangle = 1$, the overlaps between local states are:
\begin{align}
\langle \tilde{1} | \tilde{1} \rangle =
\frac{ h_x^2 + (h_z - 1)^2  }
{ h_x^2 + (h_z + 1)^2}
\equiv \nu,
\quad
\langle \tilde{1} | \tilde{0} \rangle =
\frac{2 h_x}{h_x^2 + (h_z + 1)^2}
\equiv \mu.
\label{eq:z2-basis-overlap}
\end{align}

The deformed wavefunction transforms as $|\psi(h_x, h_z)\rangle \leftrightarrow |\psi(h_z, h_x)\rangle$ under the dual transformation ${\cal D}_{\rm e-m}$.
Furthermore, in the $\mathbb{Z}_2$ case, there exists an additional sign-change duality arising from the anti-commutation relation $\{X, Z\} = 0$. Applying $\prod_l Z_l$ to the deformed wavefunction yields:
\begin{align}
\prod_l Z_l |\psi(h_x, h_z)\rangle
= P_2(-h_x, h_z) \prod_l Z_l |{\rm TC}_2\rangle
= |\psi(-h_x, h_z)\rangle.
\label{eq:z2-sign-dual}
\end{align}
Similarly, applying $\prod_l X_l$ establishes the $h_z \leftrightarrow -h_z$ duality.
Consequently, the phase diagram is symmetric about the $h_z = h_x$, $h_x = 0$, and $h_z = 0$ axes, as shown in Fig.\,\ref{fig:z2-phase-diagram}.

\subsection{Mapping to classical Ising model}

\begin{figure}[ht]
\centering
\setlength{\abovecaptionskip}{5pt}
\includegraphics[width=0.95\textwidth]{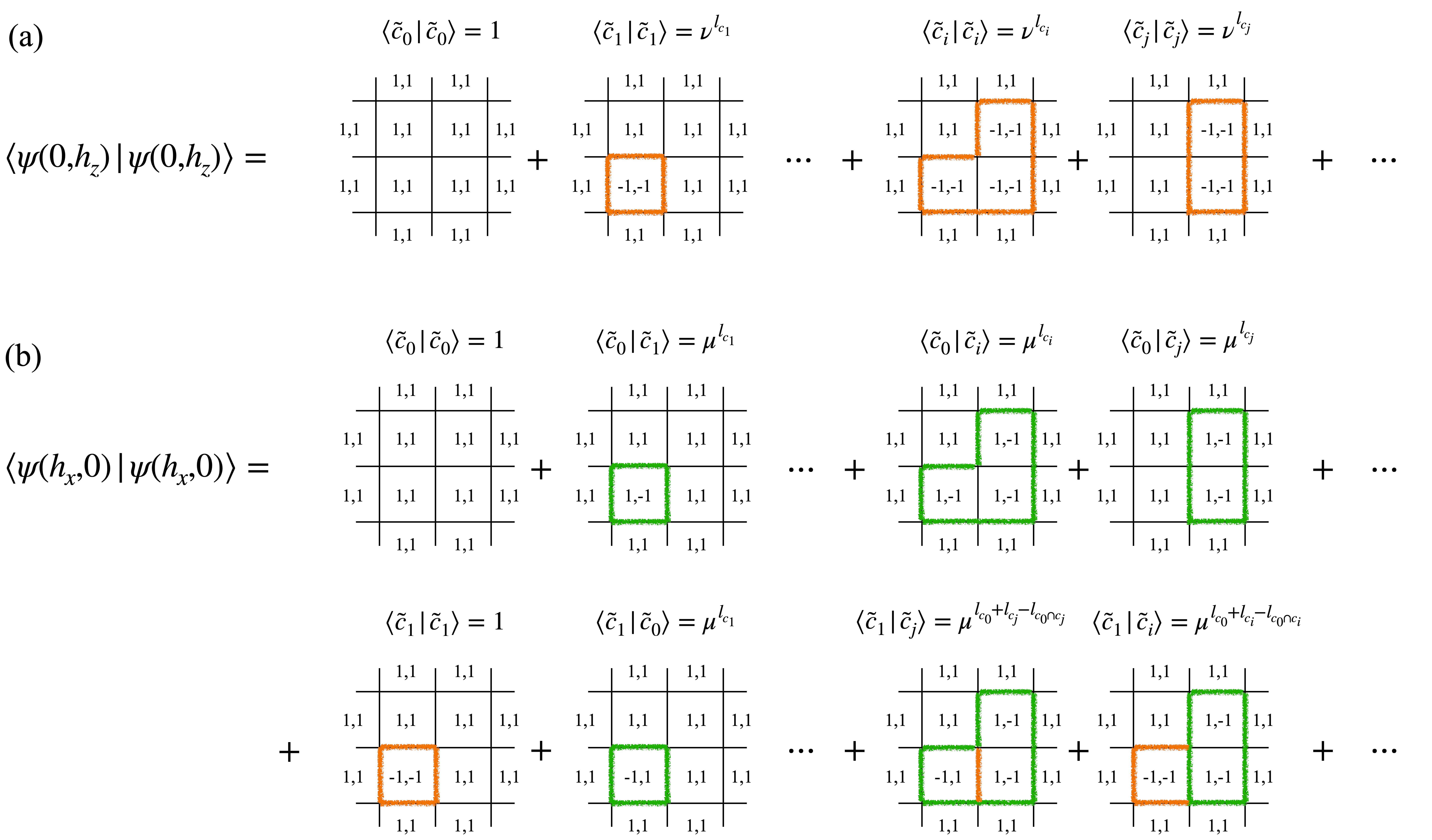}
\caption{
(a) Overlap configurations for the $h_z$ deformation ($h_x = 0$), where each configuration contributes a weight $\nu^{l_c}$. The orange loops represent domain walls of the classical Ising model.
(b) Overlap configurations for the $h_x$ deformation ($h_z = 0$), involving both orange and green loops. Green loops represent domain walls of the product spin $S_i = s_i \sigma_i$.
}
\label{fig:z2-loop-deform-1}
\end{figure}

We begin by examining the limits $h_x = 0$ and $h_z = 0$. When $h_x = 0$, the deformed local states reduce to $|\tilde{0}\rangle = |0\rangle$ and $|\tilde{1}\rangle = \sqrt{\nu}\, |1\rangle$. The norm of the deformed wavefunction becomes:
\begin{align}
\langle \psi(0, h_z)|\psi(0, h_z)\rangle
= \sum_{c} \nu^{l_c},
\label{eq:z2tc-norm-ising-1}
\end{align}
where $l_c$ is the total length of loops in configuration $c$. This is illustrated in Fig.\,\ref{fig:z2-loop-deform-1}(a), where the orange loops represent the overlap $\langle \tilde{1}|\tilde{1}\rangle = \nu$.

The norm maps to the partition function of the classical Ising model:
\begin{align}
H_{\rm Ising} = -J_1 \sum_{\langle i,j \rangle} s_i s_j,
\end{align}
with partition function $Z_{\rm Ising} = \sum_c \exp(-K_1 l_c)$, where $K_1 = J_1/T$. By identifying $\exp(-K_1) = \nu$, we find $\langle \psi(0, h_z) | \psi(0, h_z) \rangle = Z_{\rm Ising}$. The phase transition occurs at:
\begin{align}
\nu_c = \frac{1}{1+\sqrt{2}}.
\end{align}
For $\nu > \nu_c$, the orange loops proliferate and $e$ anyons are deconfined. For $\nu < \nu_c$, the loops are suppressed and $e$ anyons become confined.

On the other hand, the $h_x$ deformation with $h_z = 0$ leads to the condensation of $e$ anyons. In this case, the norms of the local states remain unchanged, but the orthogonality is broken: $\langle \tilde{0}|\tilde{1}\rangle = \mu \neq 0$.
The norm becomes:
\begin{align}
\langle \psi(h_x, 0)|\psi(h_x, 0)\rangle
= \sum_{c, c'} \mu^{l_c + l_{c'} - l_{c \cap c'}} \nu^{l_{c \cap c'}},
\label{eq:z2tc-norm-ising-2}
\end{align}
where $c$ and $c'$ represent two loop-gas configurations, as illustrated in Fig.\,\ref{fig:z2-loop-deform-1}(b). When $h_z = 0$, we have $\nu = 1$, and the norm reduces to:
\begin{align}
\langle \psi(h_x, 0)|\psi(h_x, 0)\rangle
= N_c \sum_{c'} \mu^{l_{c'}},
\end{align}
where $N_c$ is the total number of loop configurations. This maps to the partition function of a second Ising model with the product spin $S_i = s_i \sigma_i$:
\begin{align}
H_{\rm Ising'} = -J_2 \sum_{\langle i,j \rangle} s_i \sigma_i \, s_j \sigma_j,
\end{align}
with a critical point at $\mu_c = 1/(1+\sqrt{2})$. The green loops, corresponding to domain walls of $S_i$, drive the condensation of $e$ anyons: for $\mu > \mu_c$, green loops proliferate and $e$ anyons condense; for $\mu < \mu_c$, green loops are suppressed and $e$ anyons remain uncondensed.

\subsection{Mapping to Ashkin--Teller model}

In the general deformation regime with both $h_x$ and $h_z$ nonzero, both orange and green loops contribute to the norm in Eq.\,\eqref{eq:z2tc-norm-ising-2}. In this case, the overlap configurations map to the partition function of the classical Ashkin--Teller (AT) model:
\begin{align}
H_{\rm AT} = - J_2 \sum_{\langle i,j \rangle} \left( s_i s_j + \sigma_i \sigma_j \right) - J_4 \sum_{\langle i,j \rangle} s_i s_j \sigma_i \sigma_j.
\end{align}
The AT Hamiltonian combines the two Ising models described above. The three phases of the AT model \blue{---} the fully ordered phase (both $\langle s_i \rangle \neq 0$ and $\langle s_i\sigma_i \rangle \neq 0$), the partially ordered phase ($\langle s_i \rangle = 0$ but $\langle s_i\sigma_i \rangle \neq 0$), and the disordered phase ($\langle s_i \rangle = 0$ and $\langle s_i\sigma_i \rangle = 0$) \blue{---} correspond to the $e$-confined phase, the TC phase, and the $e$-condensed phase, respectively, as illustrated in Fig.\,\ref{fig:z2-phase-diagram}. This $\mathbb{Z}_2$ framework provides the foundation for the $\mathbb{Z}_3$ generalization discussed in the main text.

\end{appendix}


% Provide your bibliography here. You have two options:

%%% FIRST OPTION
% Write your entries here directly, following the example below, including:
% Author(s), Title, Journal Ref. with year in parentheses at the end, followed by the DOI number.

%\begin{thebibliography}{99}
%\bibitem{1931_Bethe_ZP_71} H. A. Bethe, {\it Zur Theorie der Metalle. i. Eigenwerte und Eigenfunktionen der linearen Atomkette}, Zeit. f{\"u}r Phys. {\bf 71}, 205 (1931), \doi{10.1007\%2FBF01341708}.
%\bibitem{arXiv:1108.2700} P. Ginsparg, {\it It was twenty years ago today... }, \url{http://arxiv.org/abs/1108.2700}.
%\end{thebibliography}

% Use your bibtex library, formatted by the SciPost style file.
\bibliography{ref-revision.bib}

%%%%%%%%%% END TODO: BIBLIOGRAPHY

\end{document}